\documentclass[]{aa}

\usepackage{float}
\usepackage{graphicx}
\usepackage{enumitem}
\usepackage{color}
\usepackage{txfonts}
\usepackage{bbm}
\usepackage[usenames,dvipsnames,svgnames,table]{xcolor}

  \makeatletter
    \renewcommand{\paragraph}{\@startsection{paragraph}{4}{\z@}%
      {-3.25ex\@plus -1ex \@minus -.2ex}%
      {1.5ex \@plus .2ex}%
      {\normalfont\small\centering}}
     
    \renewcommand{\subparagraph}{\@startsection{subparagraph}{5}{\z@}%
      {-3.25ex\@plus -1ex \@minus -.2ex}%
      {1.5ex \@plus .2ex}%
      {\normalfont\small\centering}}
    \makeatother

\newcommand{\skirt}{\texttt{SKIRT}}

\newcommand{\ramses}{\texttt{RAMSES}}
\newcommand{\music}{\texttt{MUSIC}}

\newcommand{\kms}{{ km~s$^{-1}$}}

\begin{document} 

\title{A hydrodynamical \texttt{CLONE} of the Virgo cluster} \subtitle{II. Confronting observed and synthetic galaxy population twins in a dense environment}
 
   \author{Jenny G. Sorce\inst{1,2}\fnmsep\thanks{jenny.sorce@univ-lille.fr}
          \and
          Sean L. McGee\inst{3}
          \and
          Yohan Dubois\inst{4}
          \and 
          J\'er\'emy Blaizot\inst{5}
          \and
          Alexander Knebe\inst{6,7,8}
          \and 
          Gustavo Yepes\inst{6,7}
 }
 
    \institute{Univ. Lille, CNRS, Centrale Lille, UMR 9189 CRIStAL, F-59000 Lille, France
         \and
            Universit\'e Paris-Saclay, CNRS, Institut d'Astrophysique Spatiale, 91405, Orsay, France
         \and
         University of Birmingham School of Physics and Astronomy, Edgbaston, Birmingham B15 2TT, England
         \and
            Institut d’Astrophysique de Paris, UMR 7095, Sorbonne Universit\'e, CNRS, 98 bis boulevard Arago, 75014 Paris, France
            \and
            Univ Lyon, Univ Lyon1, Ens de Lyon, CNRS, Centre de Recherche Astrophysique de Lyon UMR5574, F-69230, Saint-Genis-Laval, France
            \and 
            Departamento de F\'isica Te\'{o}rica, M\'{o}dulo 15, Facultad de Ciencias, Universidad Aut\'{o}noma de Madrid, 28049 Madrid, Spain
            \and
            Centro de Investigaci\'{o}n Avanzada en F\'isica Fundamental (CIAFF), Facultad de Ciencias, Universidad Aut\'{o}noma de Madrid, 28049 Madrid, Spain
            \and
            International Centre for Radio Astronomy Research, University of Western Australia, 35 Stirling Highway, Crawley, Western Australia 6009, Australia
            }

   \date{Received XX XX, 2025; accepted XX XX, XXXX}

%%%%%%%%%%%%%%%%%%%%%%%%%%%%%%%%%%%%%%%%%%%%%%%%%%%%%%%%%%%%%
%%%%%%%%%%%%%%%%%%%%%%%%%%%%%%%%%%%%%%%%%%%%%%%%%%%%%%%%%%%%%
%ABSTRACT%%%%%%%%%%%%%%%%%%%%%%%%%%%%%%%%%%%%%%%%%%%%%%%%%%%%%%
%%%%%%%%%%%%%%%%%%%%%%%%%%%%%%%%%%%%%%%%%%%%%%%%%%%%%%%%%%%%%

  \abstract{Galaxy clusters offer powerful laboratories for studying galaxy evolution in dense environments. {In this context,} the Constrained LOcal and Nesting Environment (\texttt{CLONE}) {project provides a} zoom-in hydrodynamical simulation of the Virgo cluster, including active galactic nucleus and supernova feedback, with a resolution down to 350~pc, {designed to mirror Virgo's observed properties. Previous work showed that this replica and Virgo} share the same history, mass, and luminosity distributions including the central M87.  {This study examines} several observational relations extending to lower stellar masses than previous synthetic-population studies: star formation density, (specific) star formation rate, metallicity, and quenched fraction of galaxies as a function of stellar mass and cluster-centric distance. {The aim is to assess how simulated and observed trends compare rather than using the replica as a stand-alone predictive model. Despite} the slightly low metallicity and the sufficiently high quenched fraction, simulated galaxies reproduce key observational trends {even without} averaging or {accounting for} observational uncertainties, aside from the consideration of projection effects: 1) At fixed stellar mass, cluster galaxies form fewer stars than field counterparts; 2) 
Most galaxies are quenched, except  for intermediate-mass or isolated galaxies; 3) Low-mass galaxies are highly quenched, thus implying a sharp metallicity drop, and low metallicity does not imply youth; 4) Quenching occurs earlier for the most massive and the smallest galaxies than for those of intermediate mass, at least until they enter the cluster; 5) Quenched galaxies have undergone dark matter stripping; 6) Gas depletion drives quenching, especially in low-mass galaxies and the farther from the cluster center they are. {Overall, the synthetic population jointly reproduces  multiple observational trends, making it a valuable} tool to probe processes from jellyfish galaxies to cluster-core gas dynamics. Upcoming runs with modified subgrid recipes will permit an exploration of their effect on metallicity and quenching rates.}
 

\titlerunning{The Virgo \texttt{CLONE} II}
\authorrunning{Sorce et al.}

\maketitle
 %\nolinenumbers
%%%%%%%%%%%%%%%%%%%%%%%%%%%%%%%%%%%%%%%%%%%%%%%%%%%%%%%%%%%%%
%%%%%%%%%%%%%%%%%%%%%%%%%%%%%%%%%%%%%%%%%%%%%%%%%%%%%%%%%%%%%
%INTRODUCTION%%%%%%%%%%%%%%%%%%%%%%%%%%%%%%%%%%%%%%%%%%%%%%%%%%%%
%%%%%%%%%%%%%%%%%%%%%%%%%%%%%%%%%%%%%%%%%%%%%%%%%%%%%%%%%%%%%
%%%%%%%%%%%%%%%%%%%%%%%%%%%%%%%%%%%%%%%%%%%%%%%%%%%%%%%%%%%%%

\section{Introduction}

Galaxy clusters constitute formidable cosmological probes provided that they are first properly understood as physics laboratories. Subsequently, an increasing number of cluster observing programs with various missions and instruments at multiple wavelengths with different purposes arose \citep[e.g.,][to name a very few]{1980ApJ...235..351U,1975A&AS...21..137J,1991A&A...252L..23P,1999A&A...345..681O,2000MNRAS.315..669F,2000ApJS..129..435B,2001A&A...365L..80A,2001ApJ...547..594R,2002ApJS..140..239C,2004A&A...423...75V,2006A&A...445..805F,2007A&A...469..363B,2007ApJS..172..182F,2007ApJS..172..561B,2012A&A...537A..39S,2014A&A...567A..65B,2014A&A...570A..69B,2016A&A...594A..32T,2018A&A...620A...1M,2018A&A...620A.198M,2019A&A...629A..14V} to increase our understanding and to develop models. The latter need to be tested via modeling   compared to observations, a necessity to refine our knowledge regarding cluster formation as well as the evolution of galaxies once they enter this dense environment. Galaxy clusters soon became indissociable from their  large-scale cosmic environment \citep{2001AAS...19910015M,2004ogci.conf...19P}. Full hydrodynamical simulations of clusters within a cosmological context grew indispensable.

However, such simulations are quite challenging due to the large dynamic range in the problem (from megaparsec to kiloparsec and even smaller scales). Multiple questions remain open regarding baryonic physics especially because of the different codes, techniques, and processes used \citep{2016MNRAS.458.4052C,2018MNRAS.480.2898C}. Until recently galaxy properties within clusters were mostly studied with semi-analytical modeling \citep[e.g.,][]{dlb07,2010MNRAS.406.1533D} or within idealized simulations \citep[e.g.,][]{2010ApJ...713.1332R}. \citet{2012MNRAS.427.1816G,2015MNRAS.447..374G} started the ball rolling using cosmological hydrodynamical simulations to reproduce a larger fraction of red galaxies within denser environments than within the field. Several studies and large projects followed at different resolutions and with different techniques \citep[e.g.,][see also \citet{2019MNRAS.483.3336T} for an extensive list of cluster re-simulations]{2014MNRAS.444.1518V,dubois14,2016MNRAS.459.4408M,2016MNRAS.463.1797D,2017MNRAS.465.2936M,2018ApJ...868..130W,2024A&A...686A.157N,2025arXiv250706301H} confirming overall the agreement with SDSS and zCosmos surveys \citep[e.g.,][]{2010AJ....139.2097P},  for instance in terms of bright cluster galaxy masses and star formation histories \citep[e.g.,][]{2017MNRAS.470.4186B,2017MNRAS.465..213B,2008MNRAS.383..593M}, baryonic content and intracluster medium properties \citep[e.g.,][]{2015MNRAS.452.1982W,2017ApJ...849...54L}, galaxy properties, and evolution \citep[e.g.,][]{2016MNRAS.458.1096E,2018MNRAS.475..648P}. 

However, since most of these successes rely on subgrid models for feedback, a complete understanding of (and a full match with) observations are still to be reached. Notably, although the modulation of star formation of galaxies within clusters (which are more likely to be quenched than in the field) is reproduced, the fraction of quenched galaxies (which depends on the large-scale environment or more generally the history of the cluster) leaves the mechanism responsible for their quenching open to question \citep[e.g.,][]{2012MNRAS.424..232W,wei12,2015ApJ...806..101H}. Several processes involve interactions with: 1) the intracluster medium via ram-pressure stripping \citep[e.g][]{2013MNRAS.429.1747M}, and strangulation or starvation because of the hot environment preventing the gas reservoir to be filled again \citep[e.g.,][]{2008MNRAS.387...79V}, or a combination of both \citep[e.g.,][]{1999MNRAS.309..161M}~; 2) other cluster members through either flyby or merger events \citep{1996Natur.379..613M,1998ApJ...495..139M} have been suggested. Additionally, \citet{2017MNRAS.470.4186B} hinted that not only does the high-density environment affect gas content and star formation rates (SFRs) of cluster galaxies, but it might also impact galaxy stellar masses out to large radii from the cluster centers, implying fundamental biases when stellar masses are used to compare similar galaxies in fields and clusters.  Complications could be added if physical processes at stakes also depend on the halo mass. The same is true for the metallicity. For instance, in order to simultaneously match the observed constraints on the intracluster medium metallicity and on the stellar mass-to-light ratios, the metal yield from supernovae must increase with the cluster mass, but in a fashion that seems unjustified by current chemical and population synthesis models \citep{2014MNRAS.444.3581R}.

Faced with these open questions, this paper uses the zoom hydrodynamical simulation of the Virgo cluster counterpart obtained with initial conditions constrained only with peculiar velocities, presented in \citet{2021MNRAS.504.2998S}, to evaluate our current understanding with a well-know galaxy cluster.  {Unlike studies relying solely on standard simulations to infer physical trends, this work adopts an observation-driven approach. The simulation is used here as a controlled numerical laboratory designed to confront a specific, well-observed cluster with the observational data available today. Our goal is to identify where the constrained simulation reproduces observed trends and where it does not, in order to delineate the regimes of validity and the limitations of our current hydrodynamical model. In this sense, the simulation provides guidance rather than definitive physical predictions. More specifically,} this zoom-in simulation of the Virgo cluster is the first one of its kind produced with the adaptative mesh refinement \ramses\ code \citep{2002A&A...385..337T}. Our first paper showed how the distribution of galaxies and the history of the observed Virgo cluster is remarkably reproduced by the simulation. This second paper in the series starts with a summary of the properties of the Constrained LOcal \& Nesting Environment (\texttt{CLONE}) simulation. In the second part it cross-checks the simulation against typical expectations for a cluster: the effect of the dense environment, the SFR, the quenched fraction of galaxies and their metallicities, and their evolution with the distance from the cluster and through the ages. It concludes with our current understanding and highlights the questions we are still faced with regarding the complex baryonic physics. Meanwhile, the Virgo \texttt{CLONE} was also used in another series of four papers to study the dynamics and structure of the intracluster medium. Hence, \citet{2025arXiv250109573L,2025arXiv250614441L}, and \citet{2024A&A...682A.157L} show that gas motions and turbulence that originate from cosmic filaments and internal sloshing, as well as projection effects, significantly impact observable properties such as X-ray and Sunyaev-Zel'dovich signals, leading to biases in cluster mass estimates. Additionally, \citet{2024A&A...689A..19L} evaluate the splashback radius as a potential boundary for clusters, but find its observability limited by projection effects and dynamical complexity.

%%%%%%%%%%%%%%%%%%%%%%%%%%%%%%%%%%%%%%%%%%%
%The simulation%%%%%%%%%%%%%%%
%%%%%%%%%%%%%%%%%%%%%%%%%%%%%%%%%%%%%%%%%%%

\section{The \texttt{CLONE} simulation}

 \citet{2021MNRAS.504.2998S} described at length the simulation. This section  solely summarizes the main concepts of  the constrained aspect, to obtain a Virgo-like cluster in the proper large-scale environment, and the hydrodynamical aspect, to obtain a galaxy population. 

\subsection{Constrained initial conditions}

The constrained simulations were designed to match the large-scale structure around the Local Group (here within a $\sim220$~Mpc sphere radius). The details of the algorithms and steps to obtain such simulations are given in  \citet{2016MNRAS.460.2015S}. The local observational data used to constrain the initial conditions of these simulations are distances of galaxies and groups \citep{2013AJ....146...86T} converted to peculiar velocities \citep{2018MNRAS.476.4362S,2016MNRAS.455.2078S} that are bias minimized \citep{2015MNRAS.450.2644S}. 

We built 200 such realizations of the initial conditions of the local Universe that all form a Virgo cluster \citep{2019MNRAS.486.3951S}. These 200 simulated counterparts of the Virgo cluster match the observations with exceeding expectations and share similar properties, including the assembly history.  Specifically, the cosmic variance is efficiently reduced down to the cluster scale \citep{2016MNRAS.460.2015S,2018MNRAS.478.5199S,2018A&A...614A.102O,2019MNRAS.486.3951S}. We selected the Virgo cluster counterpart whose properties are the closest to the average properties (radius, velocity, number of substructures, spin, velocity dispersion, concentration, center of mass offset with respect to the spherical center) of the 200-Virgos sample and a merging history in agreement with their mean history. {This Virgo \texttt{CLONE}  has a virial radius of about 2~Mpc and a virial mass of about 5$\times$10$^{14}$~M$_\odot$ \citep{2021MNRAS.504.2998S}}. To avoid periodicity problems in the local Universe-like region, the box size was set to $\sim$740~Mpc at $z=0$ \citep{2016MNRAS.455.2078S} with a $\sim$30~Mpc zoomed-in region \citep{2001ApJS..137....1B} centered on the Virgo cluster counterpart embedded with  \music~\citep{2011MNRAS.415.2101H}. This region has an effective resolution of 8192$^3$ particles for the highest level (level$=$13, dark matter particle mass resolution of $m_{\rm DM,hr}=3\times$10$^7$~M$_\odot$). Observations indicating that the environment influences some galaxy properties out to several times the virial radius \citep[e.g.,][]{2004ogci.conf..327E,2010MNRAS.404.1231V,2012MNRAS.424..232W} motivated this large zoomed-in region.

\subsection{Hydrodynamics}

A full set of key physical processes were included to form a realistic population of galaxies, following the implementation of the Horizon-AGN run \citep{dubois14,2016MNRAS.463.3948D} augmented with black hole (BH) spin-dependent feedback of active galactic nuclei (AGNs) with no further calibration tied to a cluster simulation:
\begin{enumerate}
\item Radiative gas cooling and heating assuming photo-ionization equilibrium within a homogeneous UV background from $z_{\rm reion}=10$  \citep{1996ApJ...461...20H}, including the contribution from metals released by supernovae in the cooling curve \citep{sutherland&dopita93} down to $T_{\rm min}=10^4 \, \rm K$;
\item When the gas density is greater than 0.1~H~cm$^{-3}$, random Poisson process spawning of stellar particles of mass $m_{\rm s,res}=1.4\times10^{5}\,\rm M_\odot$ \citep{rasera&teyssier06} according to a Schmidt law with an efficiency of 0.02 \citep{2007ApJ...654..304K};
\item When the gas density is greater than $0.1~\rm H~cm^{-3}$ and the local star density is greater than one-third of the local gas density, a BH is formed if there is no other BH within $50 ~\rm ckpc$;
\item Modification of the gas mass, momentum, and energy in surrounding cells of type II supernovae \citep[][updated version]{2008A&A...482L..13D}, with a release of $\sim$~10$^{51}$erg per type II supernova with a $\eta_{\rm SN}=0.2$ mass fraction of the initial mass function (IMF). Each individual stellar particle deposits $m_{\rm s,res}\eta_{\rm SN}10^{50}~\rm erg~M_\odot^{-1}$ at once after 20~Myr with a $0.1$ metal yield with respect to its own content;
\item Fraction of the rest-mass accreted energy, prescribed by the capped-at-Eddington Bondy-Hoyle-Littleton accretion rate, onto BH particles, returned into the surrounding gas to mimic the quasar-like wind release of either a radiatively efficient~\citep{shakura&sunyaev73} or radiatively inefficient accretion disk ~\citep[see][for technical details of the accretion-ejection scheme]{dubois12agnmodel}. The efficiency of the jet mode feedback is a function of the spin of the BH according to the solution of \citet[see \citealt{2021A&A...651A.109D} for more details]{mckinneyetal12}.
\end{enumerate}

\subsection{Run}

The hydrodynamical simulation was run on 5040 cores from $z=120$ to 0 with the adaptive mesh refinement \ramses\ code \citep{2002A&A...385..337T} within the Planck cosmology framework with total matter density $\Omega_{\rm m}=0.307$, dark energy density $\Omega_\Lambda=0.693$, baryonic density $\Omega_{\rm b}=0.048$, Hubble constant H$_0=67.77$\kms~Mpc$^{-1}$, and amplitude of the matter power spectrum at 8~Mpc $\sigma_8=0.829$ \citep{2014A&A...571A..16P}. 
The Euler equations are solved with the MUSCL-Hancock method. Specifically, a second-order Godunov scheme linearly interpolates, with a minmod total variation diminishing scheme, the hydrodynamical quantities at cell interfaces to solve the Euler equations with the approximate Harten-Lax-van Leer-Contact Riemann solver~\citep{Toro1994}.

The mesh in the zoomed-in region is dynamically (un-)refined from level 13 down to 21 according to a pseudo-Lagrangian criterion, i.e., when the total density in a cell is higher (lower) than the density of a cell containing eight dark matter particles at the maximum resolution would be. The initial coarse grid is thus adaptively refined down to a best-achieved spatial resolution of $\sim$350~pc roughly constant in proper length. Specifically, a new level is added at expansion factors $a=0.1,0.2,0.4,0.8$ up to level 21 after $a=0.8$.

%%%%%%%%%%%%%%%%%%%%%%%%%%%%%%%%%%%%%%%%%%%%%%%%%%%%%%%%%%%%%
%%%%%%%%%%%%%%%%%%%%%%%%%%%%%%%%%%%%%%%%%%%%%%%%%%%%%%%%%%%%%
% Cluster galaxy population%%%%%%%%%%%%%%%%%%%%%%%%%%%%%%%%%%%%%%%%
%%%%%%%%%%%%%%%%%%%%%%%%%%%%%%%%%%%%%%%%%%%%%%%%%%%%%%%%%%%%%
%%%%%%%%%%%%%%%%%%%%%%%%%%%%%%%%%%%%%%%%%%%%%%%%%%%%%%%%%%%%%

\section{Cluster galaxy population}

{More details regarding the derivation of the quantities mentioned hereafter can be found in \citet{2021MNRAS.504.2998S}. Briefly,} galaxies and dark matter halos in the zoom-in region were detected in real space using the local maxima of the star and dark matter particle density field \citep{2004MNRAS.352..376A,2009A&A...506..647T}. Merger trees gave the galaxy evolution across cosmic time. The entering time was then defined as the first time a galaxy crosses the virial radius of the cluster. Backsplash galaxies, that exited at $z=0$, represent 5\% of our total sample and would add 26\% of galaxies to the actual content of galaxies within the virial radius. Since in observations they cannot be distinguish from other galaxies, they were not considered differently as other galaxies, but future studies will be dedicated to their specificity. Star formation rates were computed over 10~Myr, 100~Myr, and 1~Gyr depending on the study. They include both in situ and ex situ formed stars, again because in observations these two populations of stars cannot be easily discriminated. Magnitudes and rest-frame colors of galaxies were derived using single stellar population models from \citet{2003MNRAS.344.1000B} and a Salpeter IMF in agreement with the hydrodynamical model used above. Hence, any other study used for comparisons was converted beforehand to a Salpeter IMF. Typically, each star particle contributes to a flux per frequency that depends on its mass, metallicity, and age. The contribution of all stars was then summed and filtered to obtain the flux in a given band. Rest-frame quantities were used and attenuation by dust was not included. Future work will look in more detail at the dust extinction effect. Galaxy metallicity and age were obtained summing over all their star particles weighted by their mass.  Following the observational papers our simulation is compared to $Z_\odot=0.02$  \citep[e.g.,][]{2005MNRAS.362...41G,2011MNRAS.416.1996R}. Finally, following observers \citep[e.g.,][]{2013A&A...556A..55I}, a galaxy is quenched if its  specific star formation rate (sSFR) derived over 100~Myr is below 10$^{-11}$~yr$^{-1}$. {We emphasize that the definition of quenched galaxies varies significantly across the literature, including fixed sSFR thresholds, offsets from the star-forming main sequence, and color-based selections. As a result, direct quantitative comparisons between different studies should be interpreted with caution. In this work, we adopt a single internally consistent definition applied uniformly to our simulated galaxy population, with the aim of assessing qualitative trends.}
 
 \subsection{Today ($z=0$): {Reduced star formation at fixed stellar mass in clusters}}

\begin{figure}[H]
 \centering 
\vspace{-1.cm}
\includegraphics[trim=1cm 1cm 1cm 1cm, clip, width=0.5 \textwidth]{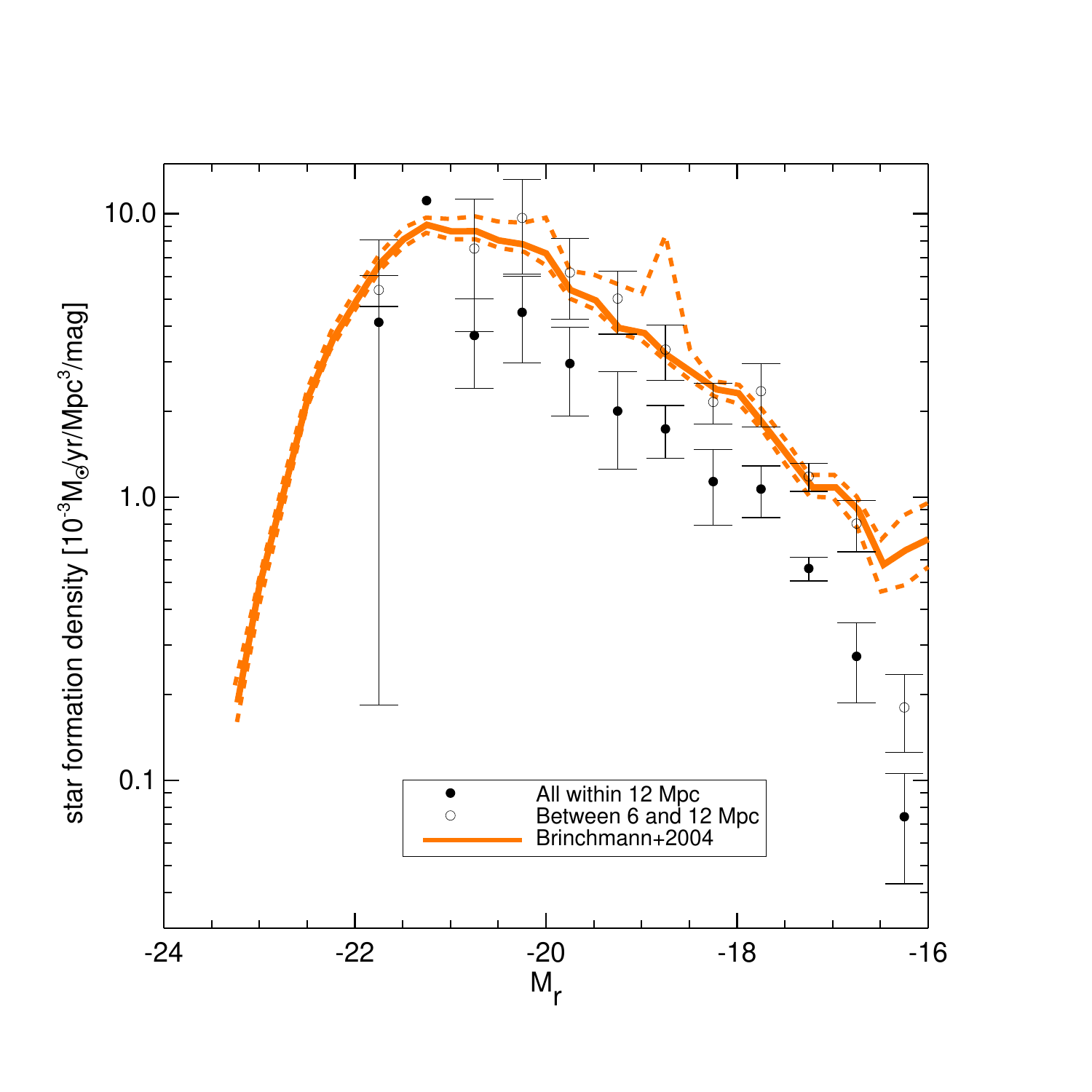}\\
\vspace{-1.5cm}

\includegraphics[trim=1cm 1cm 1cm 1cm, clip, width=0.5 \textwidth]{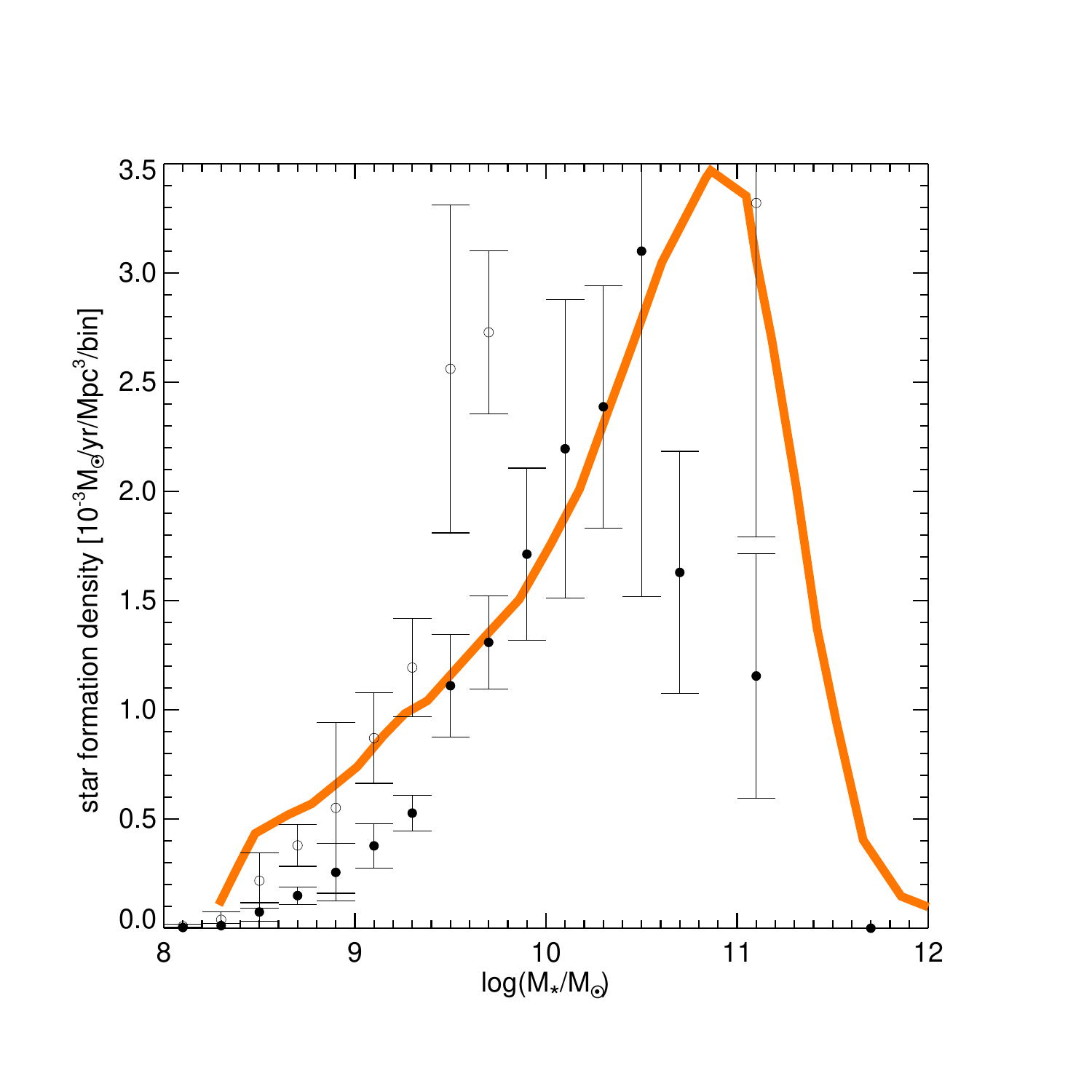}\\
\vspace{-0.2cm}

\caption{Star formation density in bins of r-band magnitude (\textit{top}) and stellar mass (\textit{bottom}). The observed star formation density shown as the solid orange line is from a volume complete down to a mass limit of 10$^{8}$~M$_\odot$ sample of SDSS galaxies biased toward field galaxies. The simulated star formation density shown as filled  black circles is obtained considering all the galaxies within a $\sim$12~Mpc radius sphere centered on the cluster, and as open black circles for galaxies between 6 and 12~Mpc from the cluster center. The dashed orange lines and black error bars in the top panel stand respectively for the uncertainties of the observational and simulated star formation density obtained with bootstrap resampling. The good agreement between the observed and synthetic relations tends to highlight the fact that galaxies in the cluster are more likely to be quenched, those of intermediate stellar mass (10$^{10}$-10$^{11}$~M$_\odot$) being somewhat spared.}
\label{fig:starformation}
\end{figure}

We first checked our simulated galaxies against general expectations obtained with SDSS galaxies in the low-redshift Universe.  Figure \ref{fig:starformation} (top) shows the star formation density of all the galaxies within $\sim$12~Mpc of the cluster center per bin of r-band magnitude as filled black circles alongside that obtained by \citet{brinchmann04}\footnote{Following their indication, their SFRs and their stellar masses have been multiplied by 1.5 to convert from a Kroupa IMF to a Salpeter IMF in agreement with our simulation, and hence goes for all observational data either from Kroupa to Salpeter or from Chabrier to Salpeter IMF. } as a solid orange line with bootstrap uncertainties as dashed lines. In our case, bootstrap resampling is shown as black error bars. Star formation histories are taken over 100~Myr.

The agreement {at about 2$\sigma$} between the observed and simulated star formation density is quite remarkable given that our simulated galaxies are in a dense environment while the SDSS sample is biased toward field galaxies. At the faintest end, the discrepancy is probably due to the mass resolution limit of the simulation around 10$^{8.5}$~M$_\odot$. However, we  also show below that although we take all the galaxies within a $\sim$12~Mpc radius around the cluster center, the faintest galaxies tend to be extremely likely to be quenched compared to observations in the field. The two aspects might thus come into play, but it is difficult to estimate which one is predominant. Regarding the brightest end, since clusters and their suburbs are dense environments, massive galaxies should be more numerous given the same volume. Thus, although   most of them are quenched, it is reasonable to think that those that are not, combined with the small statistics, contribute significantly to the star formation density and can explain the points above 10$^{-2}$~M$_\odot$~yr$^{-1}$~Mpc$^{-3}$~mag$^{-1}$ (not shown given the y-axis range). In between, galaxies in clusters and their proximity have a higher probability to be quenched, and thus their star formation density is slightly lower without an enormous difference because it is compensated by their larger number than in the field. This is confirmed by the open black circles that show the star formation density only for galaxies at more than 6~Mpc from the cluster center. The match {within less than 1$\sigma$} with the observational data is remarkable. 

The same exercise for mass bins shown in the bottom of the same figure reveals a similar trend for the star formation density in both observations and the simulation with a slight shift in the peak. This shift could be due to the uncertainty on the IMF used to derive stellar mass estimates for the observations  or to baryonic physics for the simulations. Future investigations will derive stellar mass estimates of simulated galaxies using the same methodology as that used on observed galaxies to determine how much of the shift is due to observational uncertainties, and how much is due to baryonic physics. Again, the star formation density does not appear considerably lower for intermediate-mass galaxies, although it is more peaked. The star formation density for galaxies at more than 6~Mpc from the cluster center shown with open black circles has a wider peak {confirming that the proximity with the cluster tightens the peak}. We note that it is also slightly shifted. It appears that galaxies of stellar masses between 10$^{10}$ and 10$^{11}$~M$_\odot$ are less likely to be quenched, in agreement with what is called the golden mass scale by  \citet{2025arXiv250213589T},  and this range in galaxy stellar mass reduces with the distance to the cluster. We confirm this below. For the extreme mass ends, the same applies as for magnitude bins.

In summary,  Fig. \ref{fig:starformation} shows that although our galaxies are more numerous due to their dense environment, their SFR density within 12~Mpc remains below that of the SDSS field-biased sample. This suggests that, at fixed stellar mass, they have lower SFRs, likely due to stronger quenching in the cluster dense environment.

{The top panel of} Fig. \ref{fig:stellarssfr} checks this hypothesis. It shows the average SFR of galaxies in SDSS with a yellow three-dot-dashed line alongside that of the simulated galaxies {(black filled and open circles)} within the cluster simulation. The latter are split into three groups: those within the {$\sim$2~Mpc} virial radius, called inside (in red); those within $\sim$6~Mpc, approximately the zero velocity radius,\footnote{Radius at which the mean velocity of galaxies starts to be zero from outward to inward.} but beyond the virial radius, called the outskirts (in blue); and those beyond $\sim$6~Mpc, but within $\sim$12~Mpc (to avoid the low-resolution contaminated area), called the suburbs (in black). Transparent areas stand for the 16th and 84th percentiles. Clearly at a given mass, our simulated galaxies in a cluster environment---be they inside, in the outskirts, or in the suburbs---form fewer stars by a factor of $\sim$1.4~dex than observed galaxies in the field. {The absence of a strong SFR gradient with distance from the cluster centre is not unexpected. All three regions considered here remain within the gravitational and hydrodynamical influence of Virgo and its connected large-scale structure. 

\begin{figure}[H]
\vspace{-0.9cm}

\centering
\includegraphics[width=0.45\textwidth]{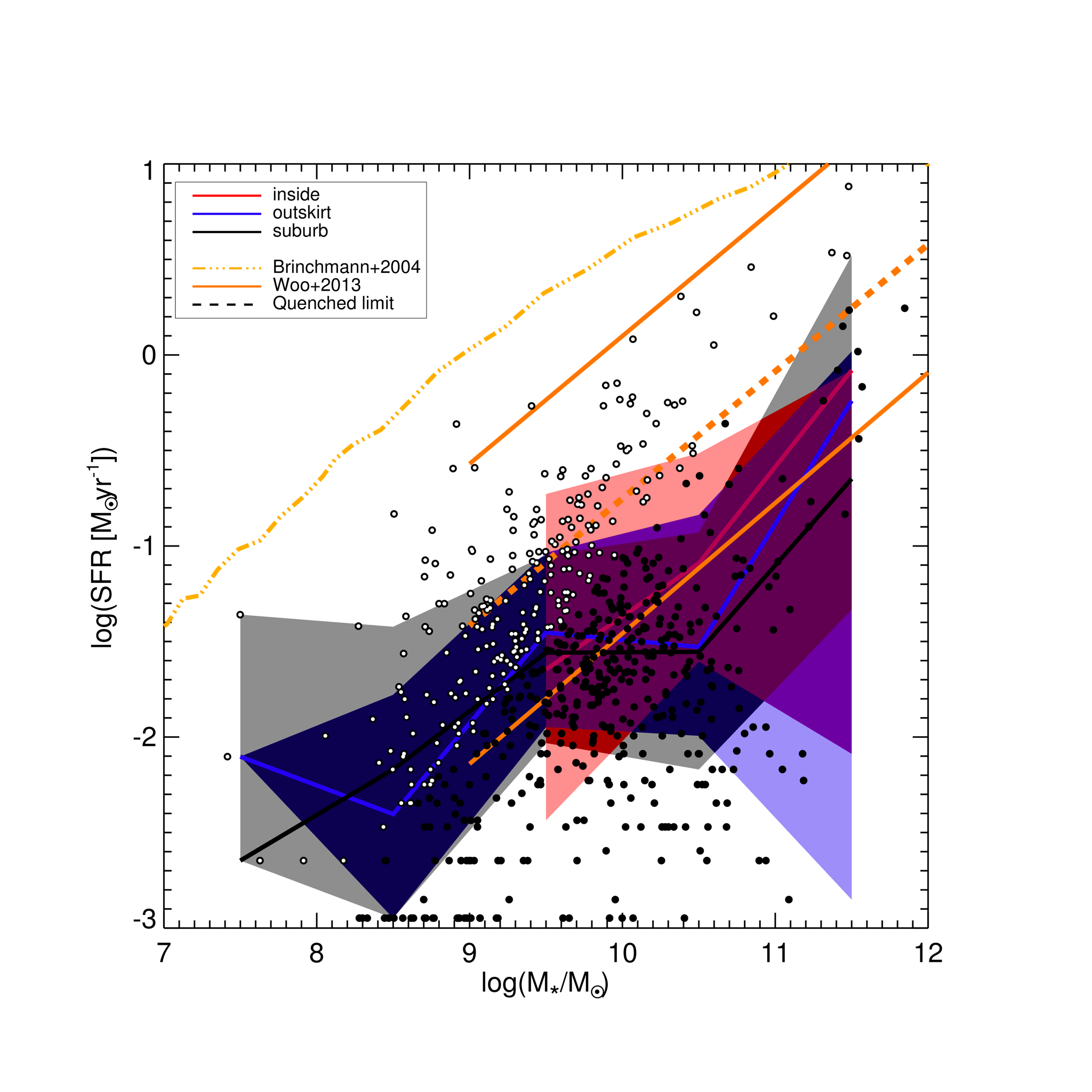}\\
\vspace{-2cm}

\includegraphics[width=0.45\textwidth]{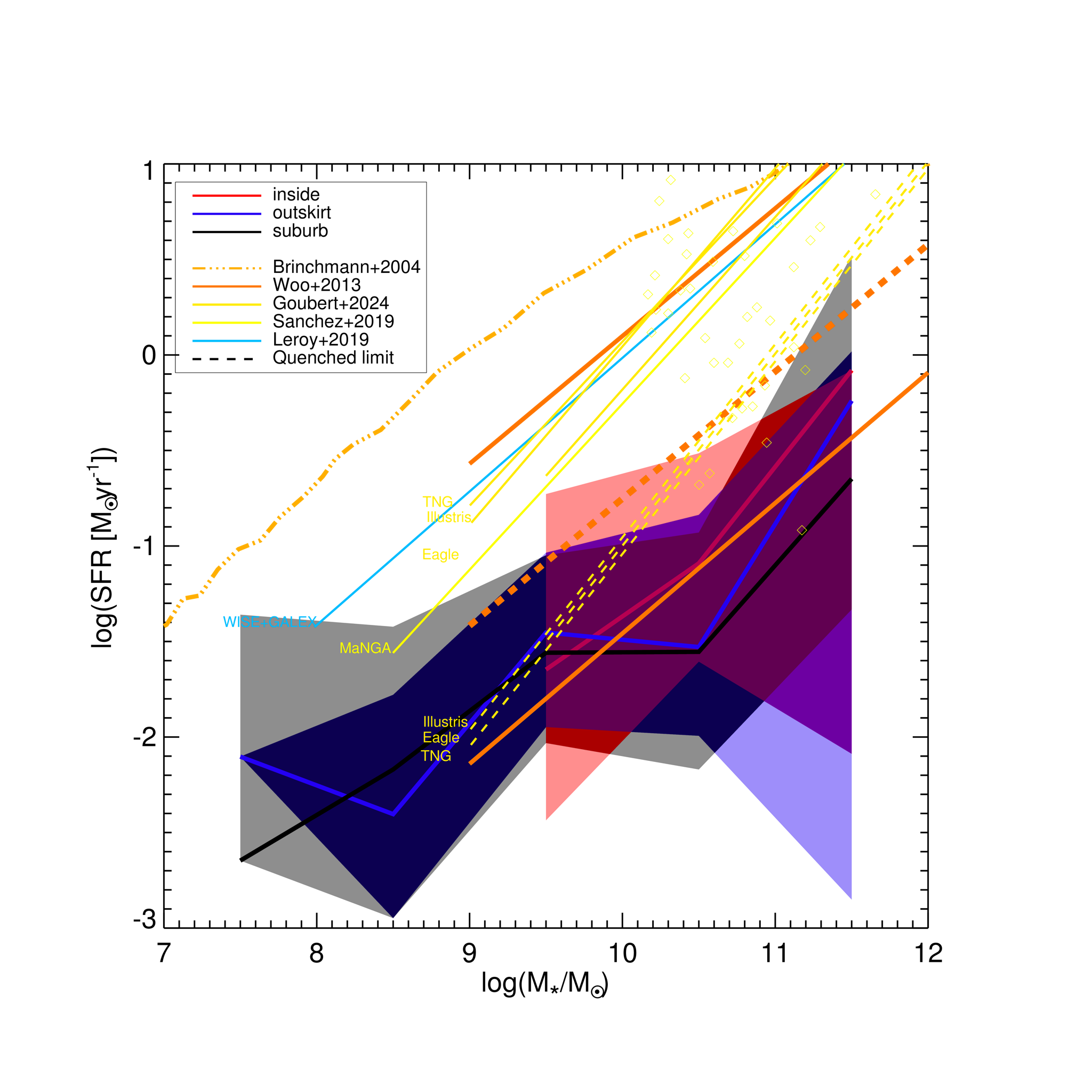}\\
\vspace{-2cm}

\includegraphics[width=0.45\textwidth]{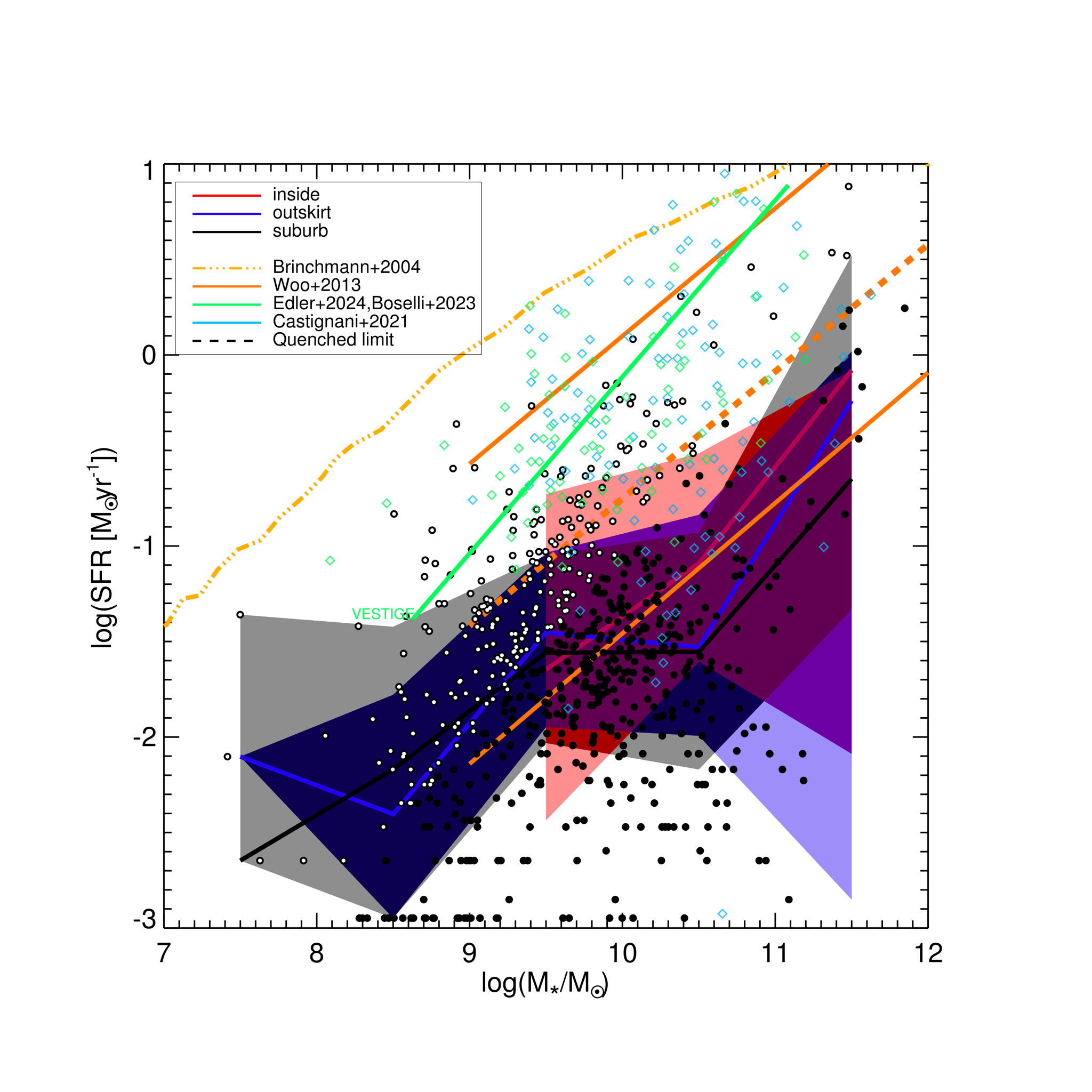}\\
\vspace{-1.cm}

\caption{Galaxy SFR as a function of their stellar mass within a SDSS complete down to a mass limit of 10$^8$~M$_\odot$ biased toward field galaxies (yellow three dot-dashed line) as well as within the virial radius of the simulated cluster (inside; red), within $\sim$6~Mpc but beyond the virial radius (outskirts; blue), and within [6,12]~Mpc (suburbs; black). The transparent areas stand for the 16th and 84th percentiles. The solid orange lines are averages obtained splitting star-forming and passive galaxies in a larger SDSS galaxy sample \citep{2013MNRAS.428.3306W}. The separation between the two regimes or quenched limit is shown as dashed orange lines in all cases. {\textit{Top}: Whole synthetic population represented as open and filled black circles if star-forming or quenched, respectively. \textit{Middle}: Synthetic populations from other simulations (light yellow lines) \citep[TNG, Illustris and Eagle,][]{2024MNRAS.528.4891G}, from a higher-$z$ sample of {environment-agnostic} galaxies called MaNGA \citep[solid light yellow line;][]{2019MNRAS.482.1557S}, and from a low-$z$ sample of {environment-agnostic}  galaxies \citep[light blue solid line,][]{2019ApJS..244...24L}. {\textit{Bottom}:} Galaxies observed within the Virgo cluster from \citet{2023A&A...669A..73B}, \citet{2024A&A...683A.149E}, and \citet{2022A&A...657A...9C} (green solid line, green open diamonds, and blue open diamonds, respectively), with those from VESTIGE biased toward inner core and ram-pressure affected galaxies.} Most simulated galaxies are quenched, meaning that the cluster-rich environment affects galaxy star formation out to large radii.} 
\label{fig:stellarssfr}
\end{figure}

Recent works have demonstrated that environmental quenching can extend well beyond the virial radius, thus affecting galaxies in filaments and infall regions up to several virial radii \citep[e.g.,][]{2017MNRAS.470.4186B}. In this context, the similar suppression of star formation across the inside, outskirts, and suburbs samples reflects the large spatial reach of cluster-driven processes rather than a shortcoming of the model. This uniformity should thus be interpreted as evidence of extended environmental quenching acting throughout the cluster’s surrounding structure.} {Moreover, the dispersion is significantly larger inside the virial radius than in the outskirts and suburbs (cf. Appendix \ref{appendixC}, Fig. \ref{fig:stellarssfrerrormedian}). This increased variance reflects small-number statistics and the presence of a few massive, recently accreted galaxies that have not yet fully quenched, and can also explain why the mean in the cluster core can appear elevated relative to the surrounding, but still cluster-affected large-scale environment.} This is therefore in agreement with the findings of \citet{2010ApJ...710L...1V}, who measure a lower median SFR by a factor of $\sim$0.2~dex for cluster galaxies than for those in the field. They warn that this measured difference for galaxies at $z<0.6$ with masses greater than 10$^{11}$~M$_\odot$ is probably a low limit because they might associate galaxies with clusters by projection. 
 
 \citet{2013MNRAS.428.3306W} extended the work of \citet{brinchmann04} with a larger sample of SDSS galaxies, but still at low $z$, and defined a quenched limit splitting galaxies between  star-forming and   passive. These two regimes are shown as solid orange lines in  Fig. \ref{fig:stellarssfr}, while the dashed line of the same color shows the limit between the two. {The lower solid orange line and the red, blue, and black lines almost overlap. The former is within the 16th and 84th percentiles of the latter}. This confirms that galaxies in our cluster simulation are mostly quenched galaxies (filled black circles) rather than star-forming galaxies (open black circles) as expected from observations. {Since determinations of the observational SFR--stellar mass relation are inherently biased toward galaxies with detectable ongoing star formation, while the simulation includes the full galaxy population, including quenched systems that may be underrepresented observationally,} their SFR-stellar mass relation matches the observed one. {Restricting the simulated sample to star-forming galaxies alone would naturally shift the SFR-stellar mass relation upward, reducing part of the apparent offset with observations.}

 \subsection{Today ($z=0$): {Flatter SFR-stellar mass relation than at $z>0$}}
 
{The middle panel of} Fig. \ref{fig:stellarssfr}  also compares our synthetic population to that obtained with other simulations \citep[TNG, Illustris, and Eagle;][]{2024MNRAS.528.4891G} in light yellow with the same line styles (solid for SF and dashed for quenched limit). Our average simulated  SFR--stellar mass relation slope {about 0.7} matches that of SDSS, contrary to the other simulations {(slopes of about 1)}. {The same is true for the zero points.} However, \citet{2024MNRAS.528.4891G} split their sample between satellite and central galaxies. Even so, the slopes obtained for the central galaxies match  the SDSS values less closely, and are not shown for clarity.  \citet{2024MNRAS.528.4891G} and  \citet{2020ApJ...902...75L} both agree that the simple central-satellite dichotomy cannot account for the difference in quenching as there is a competition between internal and environmental processes. We note though that their slope is in better agreement with the SDSS surveys such as MaNGA with a slope of {about 0.9} (solid yellow line and open diamond) that includes higher-redshift galaxies. This is probably due to a steepening of the slope with the redshift \citep{2019MNRAS.482.1557S} as again reverting to a survey of low-$z$ galaxies, for example  that of \citet{2019ApJS..244...24L}, favors slopes that are less steep   ({about 0.7;} light blue line). 
 
  \subsection{Today ($z=0$): {Stronger quenching probability of low-mass than high-mass galaxies in cluster inner cores}}
  
 Finally, {the bottom panel of} Fig. \ref{fig:stellarssfr}   shows direct comparisons with the Virgo galaxies \citep[solid green line, green open diamonds and blue open diamonds]{2023A&A...669A..73B,2024A&A...683A.149E,2022A&A...657A...9C} and reveal a general good agreement, especially with the last sample that includes galaxies up to the largest distance from Virgo core. The slope of {about 0.9} obtained for the VESTIGE sample \citep[solid green line;][]{2023A&A...669A..73B} is steeper than other relations at low $z$. However, the VESTIGE sample is  specifically targeted at galaxies inside the core of the Virgo cluster, which is in agreement with the fact that our slopes seem to increase {from about 0.5 to 0.8} from the outside  to the inside  of the cluster.  Specifically, lower stellar mass galaxies tend to be more quenched inside the cluster than outside, and  with a higher probability than higher stellar mass galaxies. The same conclusion is valid for the sample from \citet[][]{2024A&A...683A.149E}, shown as green open diamonds, that targets galaxies affected by ram-pressure stripping or close to M87. Our synthetic population most closely matches  the sample from \citet{2022A&A...657A...9C} in agreement with the fact that it gathers the largest number of galaxies of the three studies within the Virgo cluster (blue open diamonds).

In conclusion, clusters thus have   a general low star formation activity not only because their population is old and passive, but also because their star-forming galaxies produce fewer stars than their field counterparts of the same stellar mass. We note however that their star formation density, depending on the binned parameter of choice, is similar to or only slightly lower than   that of the field because they host many more galaxies per unit of volume than the field. To check that this finding is truly an effect of the cluster environment and not a bias of the hydrodynamical modeling, the result obtained for simulated galaxies in Horizon-AGN \citep[whose subgrid recipes we used,][]{2016MNRAS.463.3948D} is shown as a dotted light blue line\footnote{The same plot at $z=0.3$ for our simulated cluster galaxies does not change the conclusion. Cluster galaxies have on average barely higher sSFR (+[0.3,0.4] dex) at $z=0.3$ with respect to $z=0$ but because the formation of the cluster is also less advanced at $z=0.3$.} at $z=0.3$ {in the figure in Appendix \ref{appendixB}. Since only the sSFR is available, the latter rather than the SFR of galaxies is plotted.} The difference with \citet{brinchmann04} is only about 0.1 dex where there is no resolution limit. This confirms that the low sSFR at all masses is really a cluster environment effect. We note that  Appendix \ref{appendixC} shows the same figures, but with transparent areas representing the standard error on the median. 

 \subsection{Today ($z=0$): {Quenched galaxies experienced dark matter stripping}}

\begin{figure}[H]
\centering
\vspace{-1cm}

\includegraphics[trim=2cm 2cm 2cm 2cm, clip, width=0.45\textwidth]{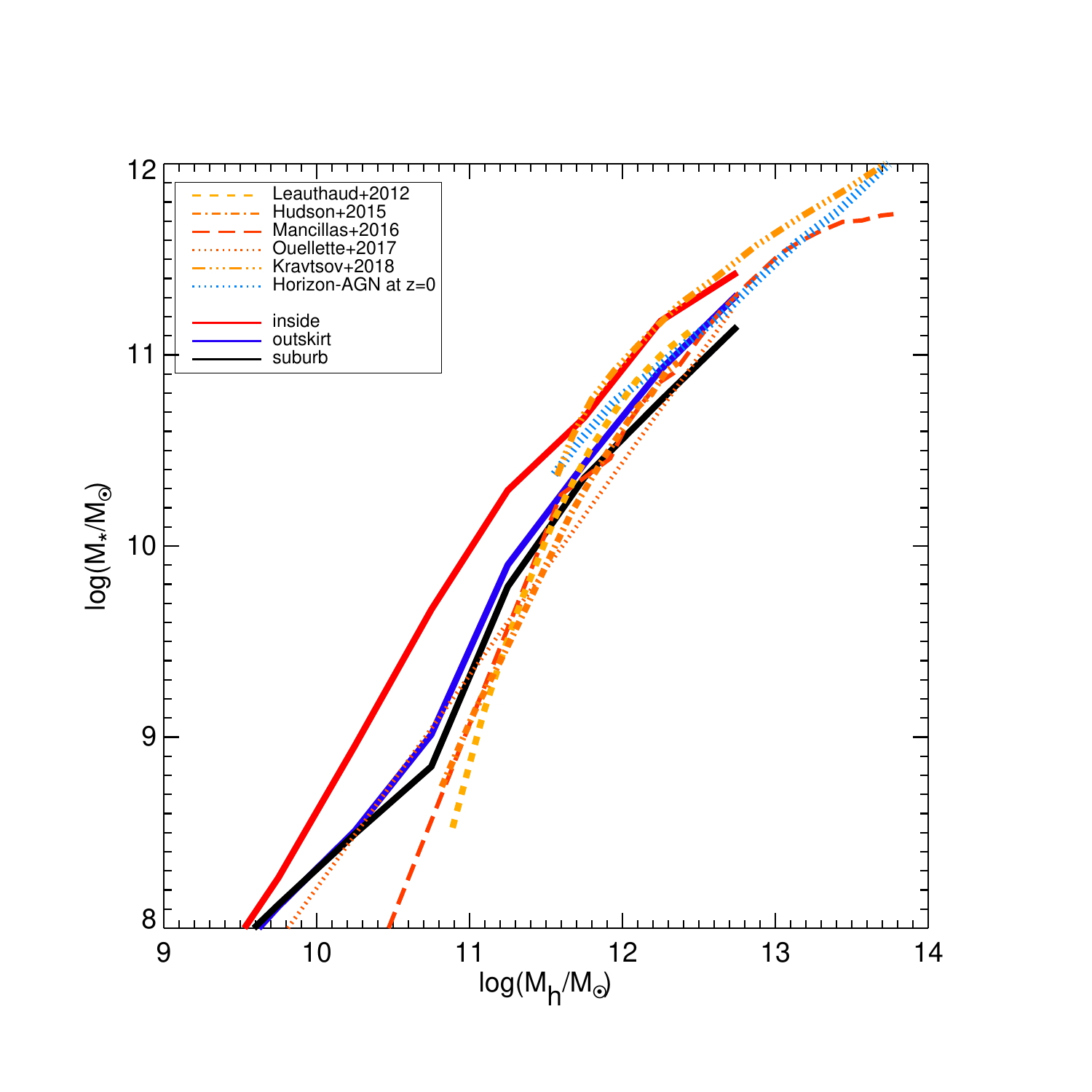} \\
\vspace{0cm}

\caption{Stellar-to-halo mass relation for galaxies inside the cluster (red), in the outskirts (blue), and in the suburbs (black) (see  Fig. \ref{fig:stellarssfr} for definitions). The relation is also given for Horizon-AGN  \citep[dotted blue line,][]{2016MNRAS.463.3948D}, for simulated late-type galaxies \citep[long-dashed orange line,][]{2017ApJ...843...74O} and for four different surveys as yellow and orange lines: galaxy clusters (three-dot-dashed), COSMOS (dashed), CFHTLenS (dot-dashed) and SHIVir (dotted; \citealt{2018AstL...44....8K,2012ApJ...744..159L,2015MNRAS.447..298H,2017ApJ...843...74O}). The agreement between the cluster survey and the simulated galaxies within the cluster is remarkable. Galaxies within the cluster have a higher stellar mass than galaxies in the field given their total mass because they have been stripped of their dark matter content upon entering in the rich cluster environment.}
\label{fig:shmr}
\end{figure}

Although galaxies in clusters have a low sSFR, their stellar-to-halo mass relation seems to show that they have formed stars in the past more abundantly than expected given their total mass today, as shown in  Fig. \ref{fig:shmr}. More precisely, according to that figure, galaxies within the virial radius in red have a higher stellar-to-halo mass relation than galaxies in the outskirts in blue {by about 0.3 dex}  than those in the suburbs in black {by about an additional 0.1 dex}. This is in agreement with observational findings. \citet[three dot-dashed line {overlapping with the red line}]{2018AstL...44....8K} use a sample of galaxy clusters, while \citet{2015MNRAS.447..298H} and \citet[dot-dashed and short-dashed lines {overlapping with the blue and black lines}]{2012ApJ...744..159L} lead their studies with a fairer sample of galaxies (CFHTLenS and COSMOS surveys). Even so, the slopes of the simulated relations might seem  a bit too shallow compared to the estimates from the observations {of the low-mass end}. It is important though to keep in mind that stellar-to-halo mass relation calibrations are very sensitive to the environmental density of the galaxies. \citet{2017ApJ...843...74O} shows that the relation for SHIVir, a representative sample of Virgo galaxies (dotted orange line), and their slope match ours. The same authors also present in the same study the Mancillas+2016 relation\footnote{We could not find any earlier, 2016 or later publication associated with this relation but keep the name given by the authors for clarity in case of any future comparison.}  obtained from a sample of simulated late-type galaxies. This last slope is in better agreement with that of the other observational studies, which implies that the slope of the relation is most probably tied with the morphological type of galaxies, and thus environment as  denser environments mean more early-type than late-type galaxies. We note that the slope obtained with Horizon-AGN, and thus the same hydrodynamical modeling, lies   between all these relations, as shown by the dotted blue line. Finally, the agreement with observational estimates in terms of shift  explains the differences between relations derived from observations. This shift is mostly due to dark matter tidal stripping \citep{2020arXiv200211119E} in the cluster environment. Specifically, galaxies had formed stars as expected given their total mass before they became partly stripped of their dark content. 

 \subsection{Today ($z=0$): {Quenching prevails, but for intermediate-mass or isolated galaxies}}
 
 \begin{figure*}
\centering
\vspace{-0.5cm}

\includegraphics[trim=1cm 1cm 1cm 1cm, clip, width=0.36 \textwidth]{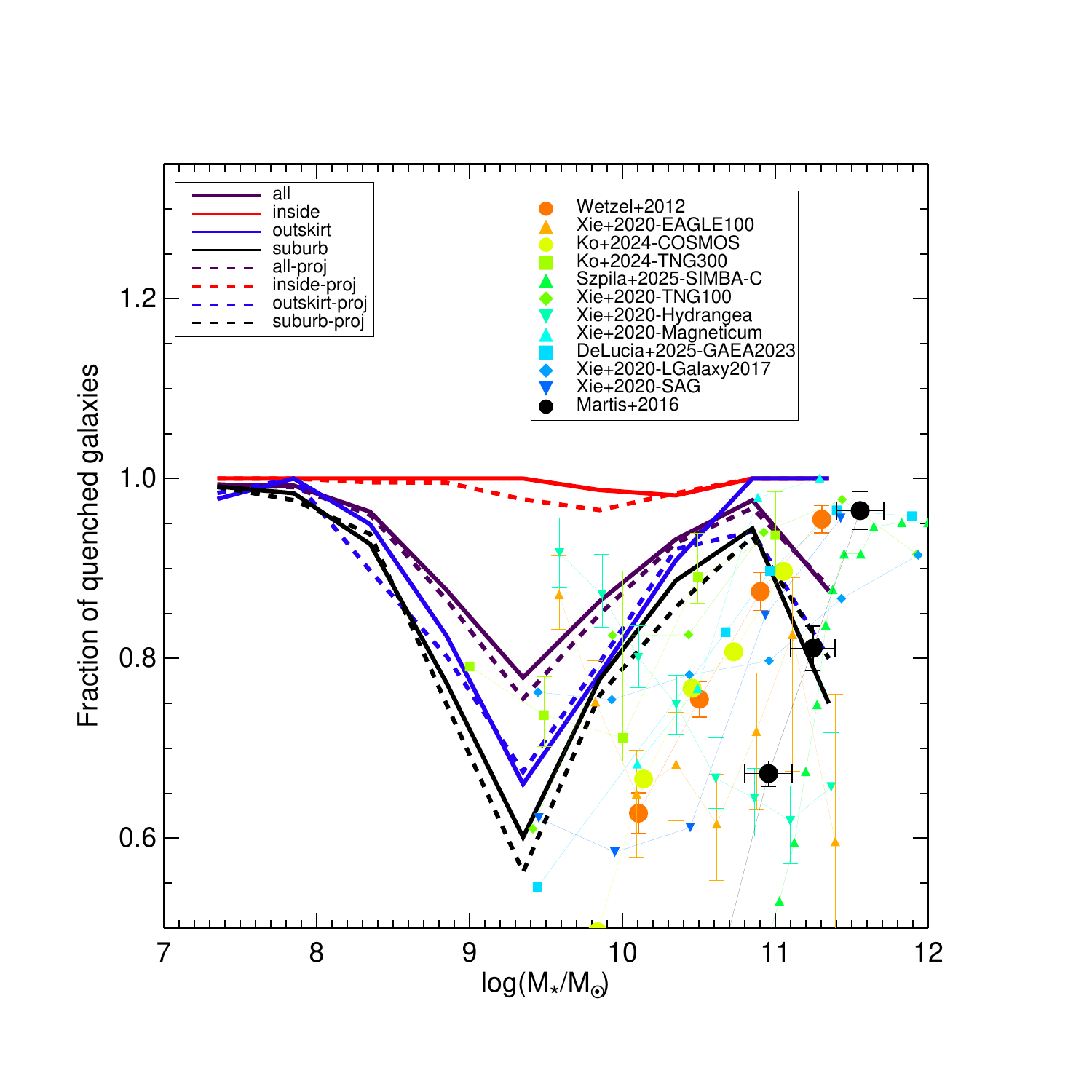} \hspace{-0.9cm}
\includegraphics[trim=1cm 1cm 1cm 1cm, clip, width=0.36 \textwidth]{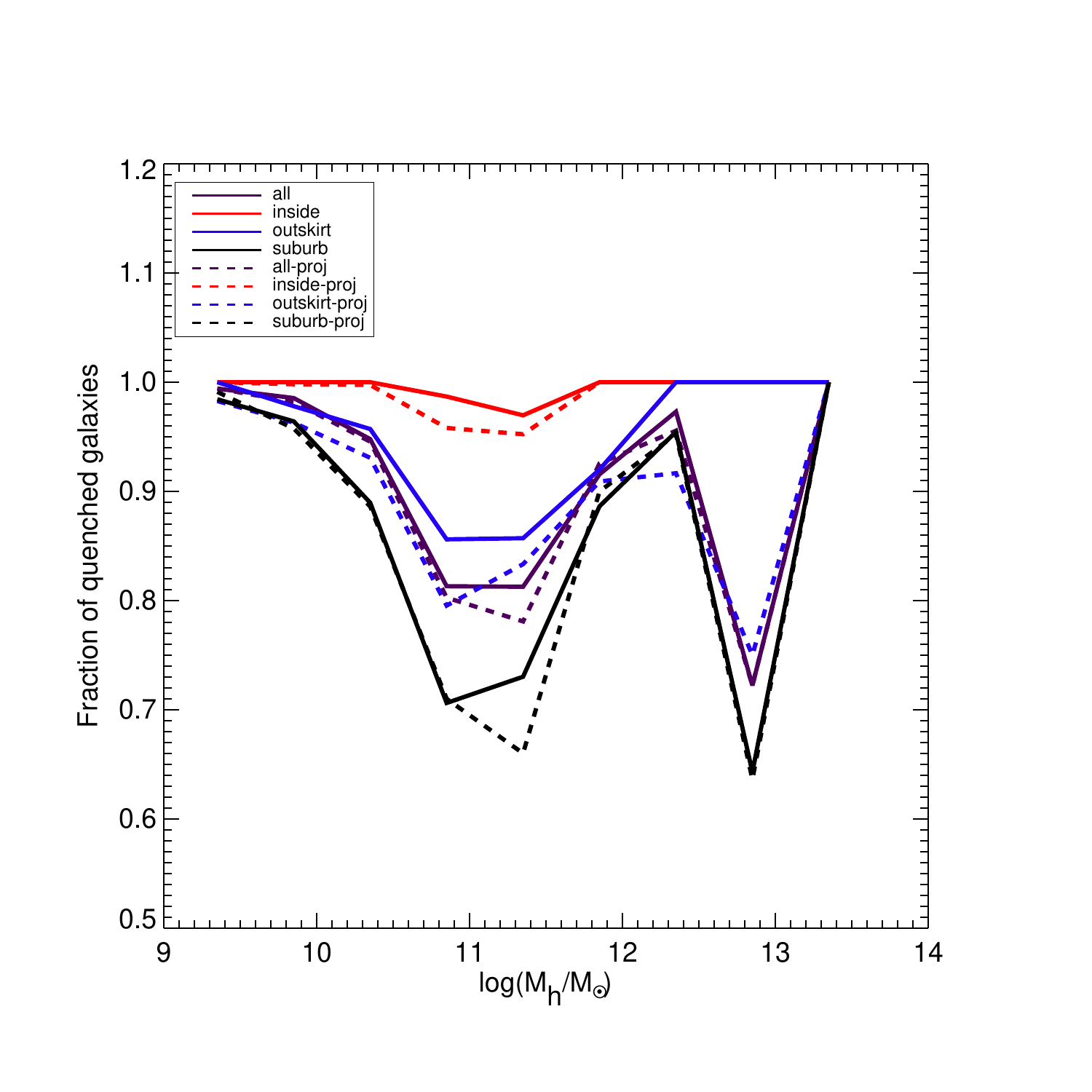} \hspace{-0.9cm}
\includegraphics[trim=1cm 1cm 1cm 1cm, clip, width=0.36 \textwidth]{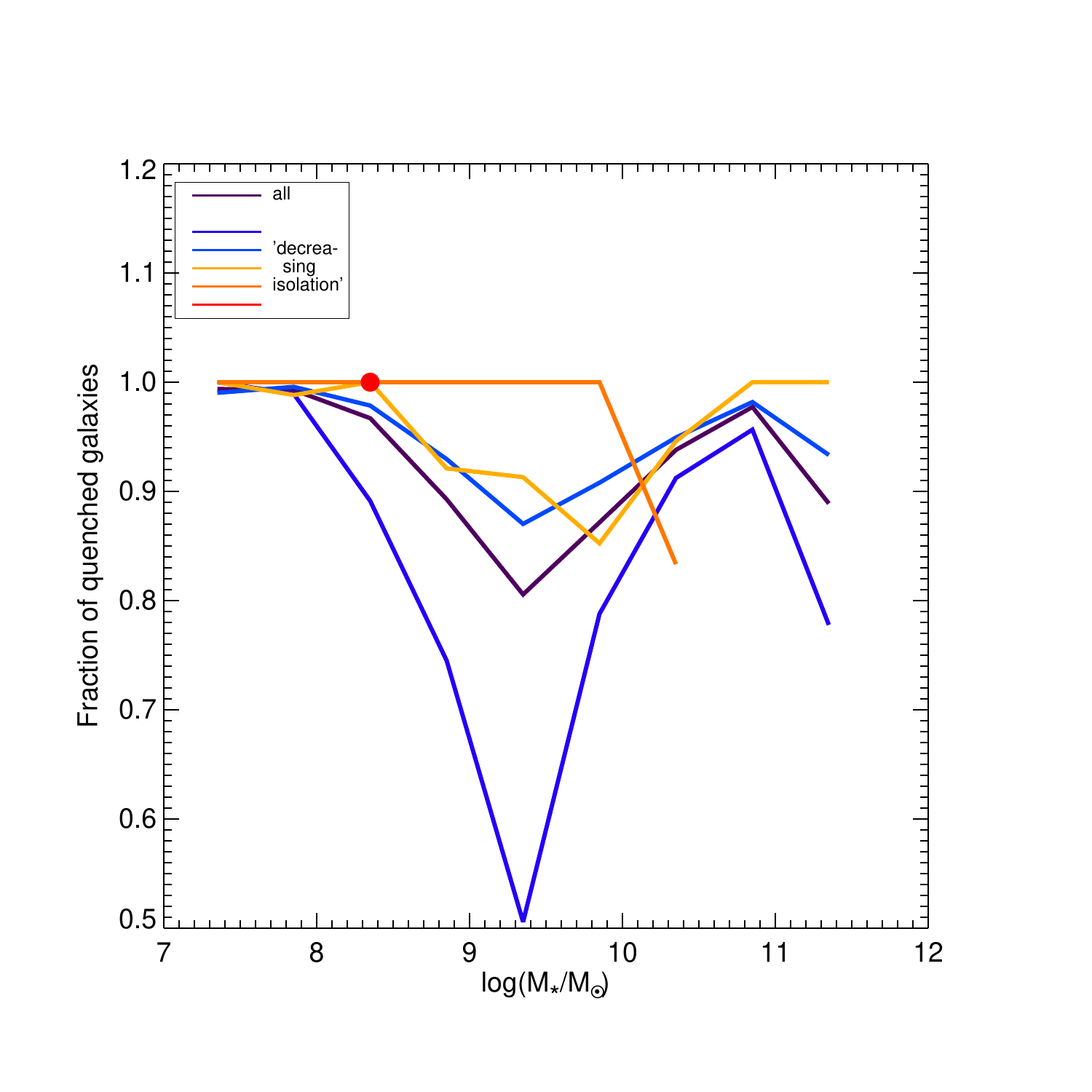}

\vspace{-0.5cm}
\caption{Fraction of quenched galaxies per bin of galaxy stellar mass (\textit{left} and \textit{right}) and galaxy dark matter halo mass (\textit{middle}). All the galaxies within a $\sim$12~Mpc radius sphere centered on the cluster are considered (violet). \textit{Left and middle}: Galaxies  split into those inside the cluster (red), in its outskirts (blue), and in its suburbs (black) (see  Fig. \ref{fig:stellarssfr} for definitions). The solid lines are derived from real galaxy distances to the cluster center, while the dashed lines are derived from projected galaxy distances to the cluster center. We note that this projection is limited by the size of the zoomed-in region. It thus constitutes only a low limit to the effect that exists in observations, all the more that the main filament linking the Virgo cluster to the cosmic web is quasi-aligned with the line of sight. \textit{Left}: Filled {symbols} and their error bars are quenched fractions and their standard deviations derived for galaxies within observed and simulated clusters and galaxy populations. \citet{2012MNRAS.424..232W} stands for the sample the closest to our cluster case as they consider observed galaxies in clusters with mass between 10$^{14.5}$ and 10$^{15}$~M$_\odot$ (large orange circles); {then galaxies from the COSMOS cluster surveys at redshift of about 0.5, but slightly lower masses between 10$^{14}$ and 10$^{14.5}$~M$_\odot$ \citep[large light yellow-green circles,][]{2024ApJ...976..154K}}; and finally  galaxies from UltraVISTA DR1 and 3D-HST {environment-agnostic} surveys \citep[large black circles,][]{2016ApJ...827L..25M}. {The smaller filled symbols stand for various synthetic  galaxy populations from simulations and semi-analytical modelings \citep[from light orange to the variety of green-blue;][]{2024ApJ...976..154K,2025MNRAS.537.1849S,2020MNRAS.498.4327X, 2025arXiv250201724D}}. Right: Galaxies   split according to their degree of {isolation}. Central galaxies are  in dark blue, while their satellites are in light blue, yellow, orange, and red. To summarize, those closest to the cluster core have the highest  probability of being quenched. The more isolated galaxies have the lowest   probability of being quenched. There exists an intermediate-mass bin where galaxies are less likely to be quenched.}
\label{fig:quenched}
\end{figure*}

We do not extend further our comparisons to colors, but rather go directly to the study of the quenched fraction of galaxies. Comparisons between observed and simulated colors are   complicated by dust contamination, and we postpone this work to future studies where codes to include dust effects such as \skirt\ \citep{2011ascl.soft09003B} will be used. We thus consider a galaxy to be passive because of its low SFR rather than because of its red color. Following observers \citep[e.g.,][]{2013A&A...556A..55I}, a galaxy is quenched if its sSFR derived over 100~Myr is below 10$^{-11}$~yr$^{-1}$.

The left panel of Fig. \ref{fig:quenched} gives the fraction of quenched galaxies as a function of their stellar mass. All the galaxies within a $\sim$12~Mpc radius sphere centered on the cluster are first considered. Then they are again split between those inside, in the outskirts, and in the suburbs. Filled {symbols} stand for quenched fraction values for observational (large circles) and simulated (small {symbols}) samples. \citet{2012MNRAS.424..232W} present results for observed SDSS galaxies in clusters of masses equivalent to our simulated cluster, our cluster being on the high-mass end of the mass range. \citet{2016ApJ...827L..25M} use galaxies from UltraVISTA DR1 and 3D-HST regardless of their environment, and {\citet{2024ApJ...976..154K} show the results for the COSMOS survey. Here we focus on cluster galaxies at redshifts of around 0.5 to minimize the evolution effect. The latter also compare their findings to results from the TNG300 simulation, which we added to the figure}. The quenched fractions from the SIMBA-C galaxy simulations from \citet{2025MNRAS.537.1849S} are also shown as well as those obtained with the GAEA2023 semi-analytical modeling \citep{2025arXiv250201724D} for their simulated clusters. {An additional collection of data points from the simulations and semi-analytical modeling are drawn from \citet{2020MNRAS.498.4327X}}. Although none of the observational and other simulated samples extend to   stellar masses that are as low as those from our simulated population, a consistent general trend can  be seen: the higher the stellar mass, the higher the quenched fraction.

{Moreover, the quenched fraction is higher for galaxies residing in cluster environments and increases with the cluster mass. At the high-mass end, the fractions\footnote{Here, the quoted quenched fractions are computed by averaging over the populated high-mass bins, rather than by selecting the single most massive bin, in order to minimize potential systematics.} reach large values, with at least 85--90\% reported for cluster surveys by \citet{2012MNRAS.424..232W} and \citet{2024ApJ...976..154K}, compared to $\sim$75\% for the environment-agnostic survey of \citet{2016ApJ...827L..25M}. Various fractions are found:  $\sim$95\% or more for cluster galaxies in TNG100 \citep[green diamonds;][]{2015A&C....13...12N}, TNG300 (green squares), Magneticum \citep[blue triangles;][]{2019MNRAS.488.5370L}, and the Virgo \texttt{CLONE} (thick dashed and solid lines) simulations; $\sim$85-90\% for the LGalaxy2017 \citep[blue diamonds;][]{2017MNRAS.469.2626H} and SAG \citep[downward blue triangles;][]{2018MNRAS.479....2C} semi-analytical models;  but only $\sim$70\% for EAGLE100 \citep[orange triangles;][]{2015MNRAS.446..521S,crain15} and $\sim$65\% for Hydrangea \citep[downward green triangles;][]{2017MNRAS.470.4186B}. We note though that the EAGLE100 values are averages obtained over halos in the mass range 10$^{14}$ to 10$^{14.5}$~M$_\odot$, i.e., slightly lower than the other samples. This could explain the slightly lower quenched fractions. However, it is not clear whether it can explain such a low value as the observational COSMOS value is also given for that mass range. GAEA2023 (blue squares) and SIMBA-C (green triangles) also have  low $\sim$70--75\% fractions, but these are environment-agnostic-like \citet{2016ApJ...827L..25M}.\\
In addition, the observations report a quite steep increase in the quenched fraction with stellar mass at the massive end. Among the synthetic galaxy populations with  high enough quenched fractions, except for the environment-agnostic ones, this sharp rise is nicely reproduced for Magneticum, TNG300, SAG, and the Virgo \texttt{CLONE}. However, the LGalaxy2017 semi-analytical model  and the TNG100 simulation do not present such a strong decline with decreasing stellar mass. 

All in all, the Virgo \texttt{CLONE} simulation reproduces quenched fractions exceeding 95\% when averaged over the highest stellar mass bins accessible within the halo, while also recovering a clear increase in the quenched fraction with stellar mass at the massive end, in qualitative agreement with observational and most of the other modelings.
This result is particularly noteworthy as most large-scale hydrodynamical studies report ensemble averages over many clusters and discard central galaxies that are less prone to being quenched, as shown in the right panel of Fig.~\ref{fig:quenched}. This means that typical studies smooth out intrinsic system-to-system variations, as shown by the error bars when available, and they increase their quenched fraction by removing  galaxies that are less likely to be quenched from their sample. In contrast, our work focuses on a single, observationally constrained Virgo-like cluster preserving all the galaxies. The fact that a single realization simultaneously reaches such high quenched fractions and reproduces the mass dependence suggests that current hydrodynamical models are capable of capturing the environmental quenching trend with mass on a cluster-by-cluster basis. At the same time, while the elevated quenched fraction may reflect genuine halo-to-halo variance, it could also hint at a discrepancy. Distinguishing between these possibilities will require a systematic comparison across several similarly constrained individual cluster simulations and their individual observations.}

{Because observers use projected distances rather than real distances to the cluster center, we also derive quenched fractions using projected distances. These distances are obtained by removing the coordinate along the line of sight, but still restricting the sample to the $\sim$24~Mpc depth size of the zoomed-in region to avoid contamination from the low-resolution area. The values derived from these projected samples therefore constitute a lower limit to the effect that could bias observational estimates. 
This lower limit is reinforced by the fact that our line of sight is almost aligned with the main filament connecting Virgo to the cosmic web, along which galaxies tend to appear more quenched \citep{2022MNRAS.517.4515S,2022A&A...658A.113M,bahe13}.  The dashed lines, obtained using projected distances, confirm that the effect tends to decrease the inferred fraction of quenched galaxies depending on the adopted line of sight. Moreover, additional observational uncertainties,  for instance in SFRs, typically not included in our simulation-based estimate, can further reduce the observed quenched fraction by up to about 20\%, as shown {by \citet{2025arXiv250315313E}}. The remaining offset thus can largely reflect the comparison between a pure single perfect 3D theoretical measurement (without projection, observational uncertainties, or sample averaging) and 2D average observational estimates, rather than a discrepancy in the underlying physics, especially given the fact that our study consists of only one cluster at the high-mass end of the range of masses considered in the other studies and at redshift zero.} 
 
Whatever sample (inside, outskirts, or suburbs) of simulated galaxies is considered, the fraction of quenched galaxies at the low stellar mass end follows a reverse trend: it decreases with the mass up to intermediate masses that are less likely to be quenched. This lower probability of being quenched for intermediate-mass galaxies is in agreement with the hint given by recent observational estimates, the GAEA2023 modeling   mentioned by \citet{2024A&A...687A..68D}, and the recent golden mass scale reported by \citet{2025arXiv250213589T}. Moreover, the lowest-mass part is in agreement with the findings of \citet{2012ApJ...757...85G}: low-mass galaxies are highly likely to be quenched when in the proximity of a cluster even outside its virial radius. We note that these results are also in agreement with those of \citet{2019MNRAS.483.3336T} for their high-resolution simulation of a random cluster but of lower mass and within a smaller zoomed-in region than ours. These galaxies might be pre-processed into other host halos as their satellites \citep{2019MNRAS.483.3336T} because of gas depletion, for example due to ram-pressure stripping \citep[e.g.,][]{2013MNRAS.429.1747M,2014A&A...570A..69B,2015ApJ...806..101H} or because of their important stellar feedback \citep{1999MNRAS.309..161M,bahe13}. We note however that the very low-mass end might be biased by the mass resolution limit, and thus by the statistics. For higher-mass galaxies, if gas depletion is the  cause, the mechanisms removing or preventing gas from flowing in need to have stronger effects. Consequently, more massive galaxies have to get closer to the halo center to become passive, and thus their quenched fraction depends more on the distance to the cluster center equivalent to our three galaxy sets: inside, outskirts, and suburbs. The split between inside--outskirts--suburbs indeed already hints at a dependence of the likeliness of a galaxy being quenched with the distance from the cluster center. A more thorough study of galaxy properties as a function of the distance center of the cluster and by bin of stellar mass is conducted in Appendix \ref{appendixA}. 

The dependence of the quenched fraction on the galaxy dark matter mass is similar to that on the stellar mass, as shown in the middle panel of  Fig. \ref{fig:quenched}. The smaller fraction of quenched galaxies at high dark matter halo mass ($\sim$75\%) than at high stellar mass ($\sim$95\%) is most probably the result of dark matter tidal stripping that happened to galaxies that are now quenched: galaxies of higher dark matter mass today appear  less likely to be quenched and a fortiori then less likely to have undergone tidal stripping of their dark matter, and thus they preserve their dark matter mass. Instead, those of slightly lower dark matter mass today are more likely to be quenched and consequently more likely to have had their dark matter stripped. It also explains why rather than a peak, as for the intermediate stellar mass, there is more a minimum plateau at intermediate dark matter halo mass. Galaxies quenched and thus stripped of their dark matter move to smaller dark matter halo mass bins.

Finally, the right panel of Fig. \ref{fig:quenched} shows that isolated\footnote{Fully isolated galaxies reside in their own independent dark matter halos, while satellite galaxies reside in subhalos embedded within a more massive host halo that contains other galaxies.} galaxies are less likely to be quenched than satellite galaxies. This panel  clearly shows that the more {isolated} a galaxy is, the less likely it is to be quenched, whatever its stellar mass, with a minimum between 10$^9$ and 10$^{10}$~M$_\odot$. Satellite galaxies are thus simply more likely to be quenched, in agreement with observations. In addition, high stellar mass galaxies are more likely to be quenched already as central galaxies, and thus the red sequence in this mass regime is not built up by satellite-specific processes as much as in the intermediate-mass regime.

 \subsection{Today ($z=0$): {Pervasive gas depletion of low-mass galaxies up to three virial radii}}
 
\begin{figure}[H]
  \centering
  \vspace{-1cm}
  
  \includegraphics[width=0.5\textwidth]{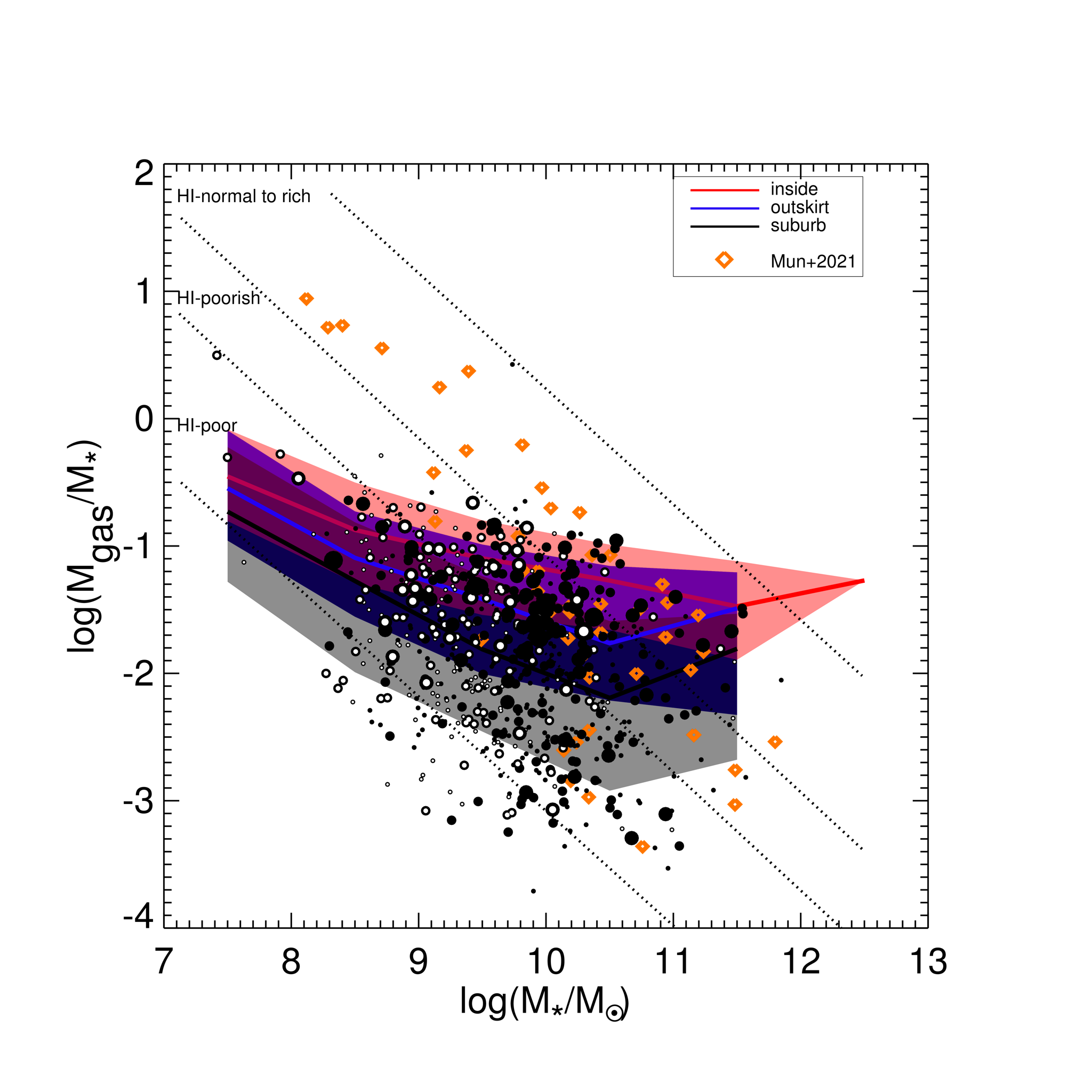}\\
\vspace{-0.2cm}

\caption{Ratio of the gas mass to the stellar mass as a function of the stellar mass of the galaxies. Simulated galaxies are split into galaxies inside the cluster (red), in its outskirts (blue), and in its suburbs (black) (see  Fig. \ref{fig:stellarssfr} for definitions). The dotted black lines delimit the regions that include galaxies poor in gas to those normal or rich in gas in the Virgo cluster, according to \citet{2021JKAS...54...17M}. In addition, the orange diamonds show the galaxies they select from their sample that are not impacted to those highly impacted by ram-pressure stripping (from top left to bottom right in the diagram). The open and filled black circles stand for our star-forming and quenched galaxies, respectively. The size of the circles is proportional to the {isolation} of the galaxy: the less isolated the galaxy, the larger the circle. Interestingly, at fixed stellar mass, galaxies in the suburbs are on average more gas poor than those in the outskirts, which are themselves more gas poor than those inside the cluster, thus suggesting that low-mass galaxies are indeed pre-processed via gas depletion before entering the cluster.}
\label{fig:mgasmstar}
\end{figure}

\citet{2010AJ....139.2097P} demonstrated that, at least at low $z$, the fraction of quenched galaxies depends on  stellar mass, because of internal processes such as stellar and AGN feedback \citep[e.g.,][]{2014MNRAS.441..599B} or morphological, gravitational, dynamical, bar, and/or angular-momentum quenching \citep[e.g.,][respectively]{2009ApJ...707..250M,2014ApJ...785...75G,2020MNRAS.495..199G,2015A&A...580A.116G,2020MNRAS.491L..51P}, and on environment, because of gas depletion via tidal or ram-pressure stripping \citep[e.g.,][]{1972ApJ...176....1G,2000Sci...288.1617Q,2018ApJ...857...71B},  strangulation and/or starvation \citep[e.g.,][]{2000ApJ...540..113B,2008MNRAS.383..593M,2015Natur.521..192P},  gravitational interactions such as galaxy-galaxy mergers \citep[e.g.,][]{1994ApJ...425L..13M,2014MNRAS.440..889S}, harassment \citep[e.g.,][]{1981ApJ...243...32F,1999MNRAS.304..465M,2014MNRAS.444.2938H}, or even pre-processing before entering the cluster in the case of a cluster environment \citep[e.g.,][]{2004cgpc.symp..277M,2012MNRAS.423.1277D,2018ApJ...866...78H,2018ApJ...865..156J}. To check this assertion and understand better the mechanisms responsible for quenching as a function of the stellar mass or environment,  Fig. \ref{fig:mgasmstar} shows the ratio of the gas mass\footnote{In the simulation, a galaxy gas mass is defined as the sum of the gas in all the cells within a two times r$_{200}$ side cube centered on the galaxy. Tests varying slightly the size of the cubes did not affect drastically the galaxy gas masses hence derived.} to the stellar mass of galaxies as a function of their stellar mass, be they inside the cluster (red), in its outskirts (blue), or in the suburbs (black). {The dotted black lines split the diagram into three different zones depending on the gas content of galaxies, following the classification of \citet{2021JKAS...54...17M}. Additionally, a subset of galaxies from the Virgo sample of \citet{2021JKAS...54...17M} is shown as orange open diamonds, ranging from those least affected by ram-pressure stripping (top left, HI-normal to rich region) to those exhibiting the most prominent signs of stripping effects (bottom right, HI-poor area). It is important to note, however, that the observational sample of \citet{2021JKAS...54...17M} is not a complete census of Virgo galaxies, but rather a targeted selection of systems chosen to span the full range of ram-pressure stripping signatures. Consequently, the observed HI fractions in their study are biased toward galaxies with unusual gas properties and do not reflect the overall Virgo population.  In contrast, our simulation includes all galaxies within the selected volume, independent of morphology or gas content, naturally producing a broader and more uniform distribution of gas fractions. Moreover, the gas masses plotted from the simulation represent the total gas content as measured within the chosen aperture, rather than an observationally corrected HI mass. Converting these total gas masses to an observational HI estimate would require a separate post-processing step, which is beyond the scope of the present paper as our goal is only to compare trends and regions. Therefore, differences between the observed and simulated trends should be interpreted primarily as arising from sample-selection effects and methodological differences, rather than as discrepancies.}
 
 In the Virgo cluster environment, low-mass galaxies tend to be gas deprived whatever their distance to the cluster center is, in agreement with the fact that strangulation may operate out to three virial radii \citep[e.g.,][]{bahe13,2018MNRAS.475.3654Z} and may also occur in small groups in a pre-processing form \citep[e.g.,][]{2009MNRAS.399.2221B,2008ApJ...672L.103K}. However, the size of the black filled (quenched) and open (star-forming) circles is proportional to galaxy's {isolation}: the larger the circle, the less isolated the galaxies, which can thus be pre-processed. Hence, we note that at a fixed stellar mass,  the most gas-deprived galaxies are not necessarily those in groups (not the largest circles), and thus gas depletion in this case is not necessarily the main mechanism of pre-processing. Moreover, these galaxies are not necessarily quenched (open rather than filled circles). Consequently, strangulation that occurs on larger timescale is probably the preponderant mechanism for gas depletion in the suburbs. Specifically,  galaxies are gas deprived by strangulation, but their quenching happens on longer timescale, and thus they are still star-forming. Closer to the core, however, ram-pressure stripping is  certainly at play \citep[e.g.,][]{2021A&A...646A.139B} although it may also operate farther out \citep[e.g.,][]{2020MNRAS.498.4327X}, including in galaxy groups \citep[e.g.,][]{2013MNRAS.436...34C}, in particular in the outskirts where the circles are more filled and larger on average. In addition, the more massive the galaxies, the more gas rich they are, which suggests that gas depletion is not the main reason for massive galaxy quenching. Internal mechanisms, for instance related to AGN feedback, are more likely.

{Interestingly, galaxies located farther from the cluster center, in the suburbs, exhibit the lowest gas mass-to-stellar mass ratios, regardless of their stellar mass. This indicates that mechanisms responsible for gas depletion can be more efficient, or at least more easily decoupled from other processes, in the cluster suburbs compared to the core. Consequently, in our simulation the apparent decrease in HI content with cluster-centric distance likely reflects environmental pre-processing in infalling groups and filaments rather than a deficiency in the physical modeling. A detailed analysis of the specific mechanisms affecting galaxy gas content and the relative contributions of internal versus external processes at different redshifts is beyond the scope of this work and will be addressed in a dedicated future study, as it warrants a full investigation on its own.}

 \subsection{Today ($z=0$): {Low metallicity in low-mass galaxies reflects quenching, not youth}}

\begin{figure}[H]
  \centering
  \vspace{-0.1cm}
  
\includegraphics[width=0.45 \textwidth]{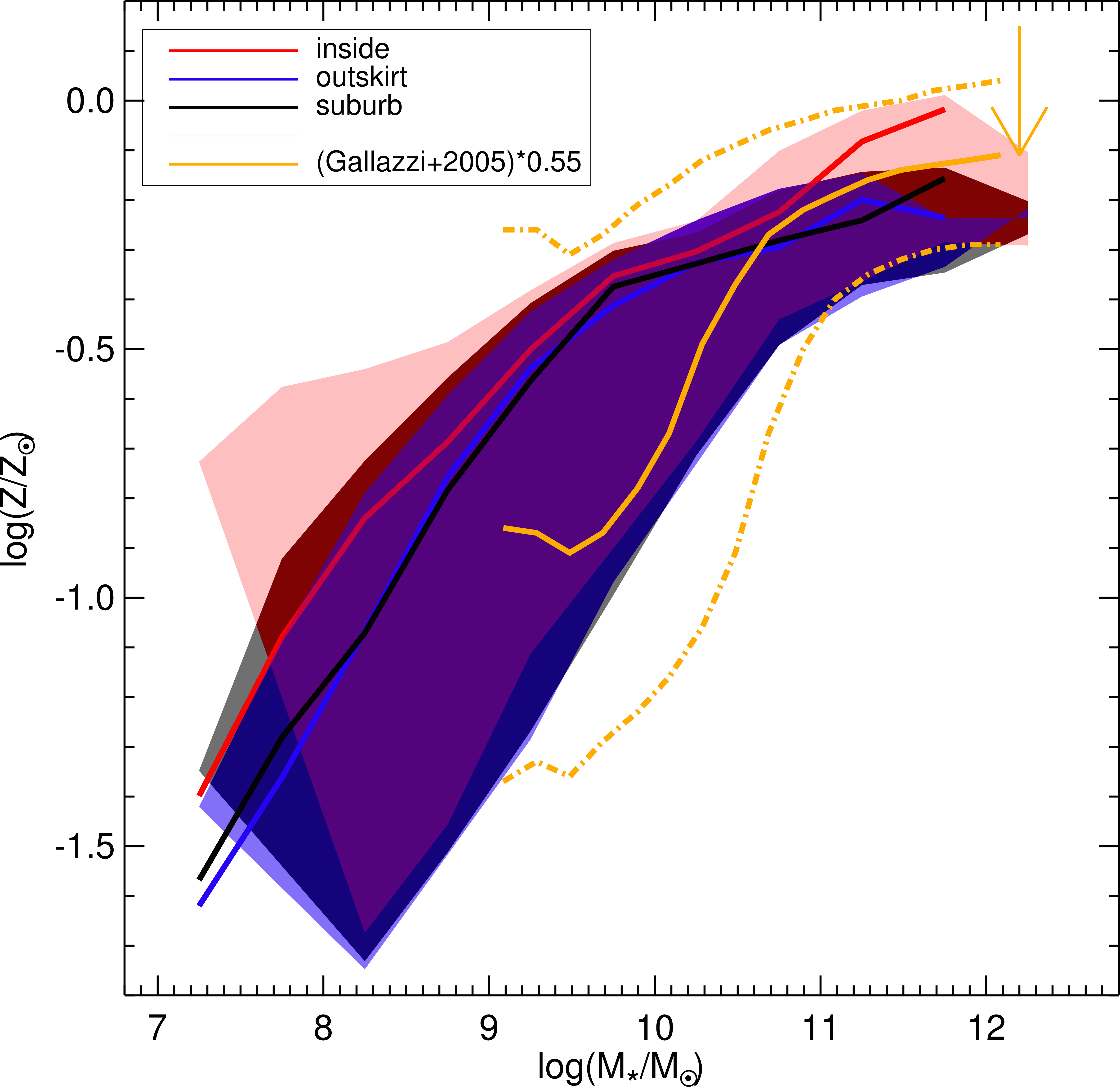}\\
\vspace{-0.2cm}

\caption{Galaxy metallicity as a function of  stellar mass. Simulated galaxies are split into galaxies inside the cluster (red), in its outskirts (blue), and in its suburbs (black) (see  Fig. \ref{fig:stellarssfr} for definitions). Metallicities of a magnitude-limited sample of SDSS galaxies in the local Universe are shown in yellow. The solid lines stand for median values per mass bin. The transparent areas as well as the dot-dashed lines represent the 16th and 84th percentiles. The yellow arrow indicates the amount by which the observed metallicities have been shifted to match the simulated values. The simulated stellar mass--metallicity relation trend and scatter very closely match the observed relations. Overall, the more massive the galaxy, the higher its metallicity.}
\label{fig:stellarmetal}
\end{figure}

\begin{figure*}
  \centering 
  
\includegraphics[width=1\textwidth]{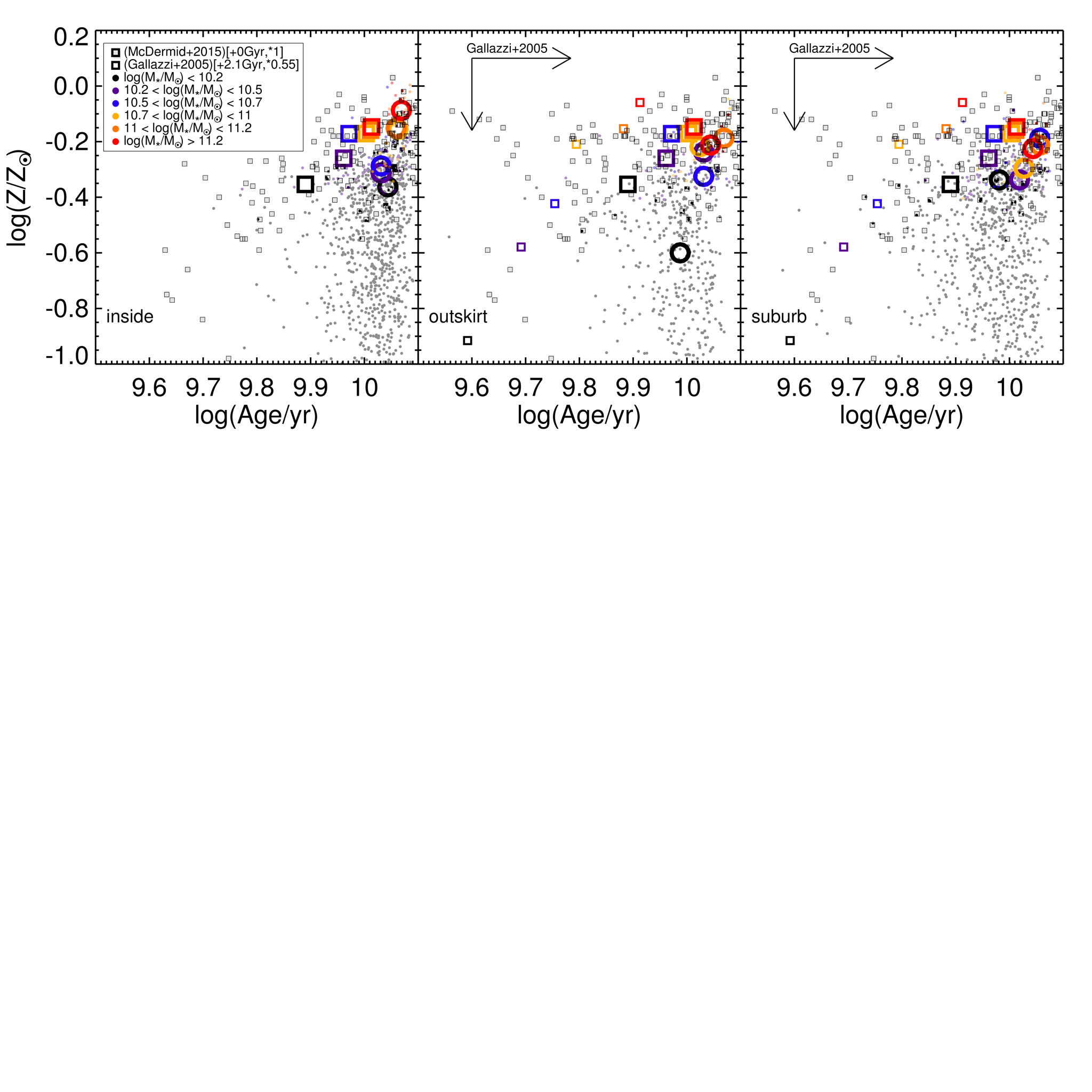}\\

\vspace{-11cm}
\caption{Galaxy metallicity as a function of age. The small colored filled circles stand for individual simulated galaxies, while the large colored open circles stand for their mean per mass bin. The colored open large (resp. small) squares represent the average observed values obtained with SDSS early-type (resp. late-type) galaxies using the same mass bin. The arrows indicate the shift applied to the observed metallicities and ages of \citet{2005MNRAS.362...41G} to match the simulated values and because the simulated galaxies belong to a cluster, respectively. The light gray filled squares are values of early-type galaxies from Altas$^{3D}$. We note that no shift has been applied  in this case since some of these individual galaxies are within the Virgo cluster and the metallicity is measured within the effective radius. Unlike for field galaxies, a low metallicity is not as much a sign of youth in the cluster. However, such a trend is more pronounced for galaxies in the suburbs of the cluster than inside its virial radius.}
\label{fig:metalage}
\end{figure*}

 Figure \ref{fig:stellarmetal} shows galaxy metallicities as a function of their stellar mass.  Simulated galaxies are again split into three samples: inside, outskirts, and suburbs. The trends are similar for the three groups: low-mass galaxies are metal poor, while massive galaxies are more metal rich. There is a noticeable break in the metallicity--stellar mass relation with a shallower slope at stellar masses above 10$^{9.5}$~M$_\odot$ than at lower stellar masses. The solid yellow line gives the same relation for a magnitude-limited sample of SDSS galaxies \citep{2005MNRAS.362...41G}.\footnote{They do not explicitly mention the IMF they use. However, since they compare their stellar mass estimates to those of \citet{2003MNRAS.341...33K} who use a Kroupa IMF, we apply the conversion for the latter.} Beforehand, we multiply the observed metallicity values by a factor 0.55 ($\sim$0.26~dex). It has been shown  that AMR \ramses\ simulations so far produce galaxies with low metallicities \citep[e.g.,][]{2011MNRAS.417.1853D,2016MNRAS.459.4408M,2017MNRAS.470..166H}. Several reasons might be at stake in addition to the metal yield based on a Salpeter IMF in our simulation. This will be the object of further investigations in a future paper with a second simulation run based on a Kroupa IMF metal yield. We can also mention that Eulerian schemes diffuse metals at the resolution scale.  \citet{2021MNRAS.504.2998S} showed that there is a gradient in metallicity from inward to outward by $\sim$0.4~dex in the M87 counterpart that makes it more metal rich when considered globally than M87 in its inner part. In a future paper, we will also show that the metallicity is quite low in the intracluster medium. Thus, the diffusion outside galaxies might not be the reason for the difference observed here. On the observational side though, systematics might affect the estimates because of i) the choice of the mix of star formation history model by about +0.04~dex, ii) the S/N cutoff by about +0.2~dex for M$_*<~$2$\times$10$^9$~M$_\odot$, or iii) the small aperture radius for SDSS spectra  \citep{2005MNRAS.362...41G}. Combined with a mass return and metal yield from a Kroupa IMF, these systematics might close, or at least reduce, the 0.26 dex difference. We also note  that  Fig. \ref{fig:metalage} shows that no shift is applied, because it was not significant, for galaxies from Altas$^{3D}$, a sample containing Virgo cluster galaxies \citep{2015MNRAS.448.3484M}, as detailed below. 

All these parameters affect the zero point, but not the trend of the relation. Thus, the agreement with the shifted observed relation is remarkable, in the sense that the observed break in the slope and scatter are both recovered. The scatter is far larger than the observational uncertainties, and thus must be reproduced by the simulation. Part of the scatter is most probably due to galaxy morphology as metallicity and age are not solely mass related \citep{2005MNRAS.362...41G}. For the transition mass, \citet{2013ApJ...772..119L} interpret the origin of the mass--metallicity relation as follows. The ratio of sSFR to gas accretion rate that changes with stellar mass determines the mass--metallicity relation. Consequently, the stellar mass threshold beyond which most galaxies are quenched is that when the slope becomes shallow. From  Fig. \ref{fig:quenched}, it indeed corresponds to the mass range with the minimal quenched fraction of galaxies. We also find back the small shift between the simulated stellar mass range where the change in slope occur and the observed one. 

Galaxy metallicities as a function of their age are shown in  Fig. \ref{fig:metalage} as colored filled circles for individual galaxies and colored open large circles for their mean. Each color represents a different mass bin. From red to black, galaxies are less massive. Overall, the youngest galaxies tend to be metal poor, while the oldest are the most metal rich. The most massive galaxies are also the oldest and the most metal rich, while the smallest galaxies are youngest and the most metal poor. We note however that depending on the galaxy position inside or outside the cluster, the trends are more or less pronounced. From left to right, galaxies are inside, in the outskirts, and in the suburbs of the cluster, and  the slope becomes shallower from inward to outward. Additionally, galaxies inside are on average older than those outside. However, their metallicity values are comparable, except for the extreme mass ends. These findings are in agreement with several studies that report an age difference between early-type galaxies in clusters and outside of about 1.2 up to 2-3 Gyr when compared with galaxies in the field \citep[e.g.,][]{1998ApJ...508L.143B,2002MNRAS.337..172K}.

Colored open squares stand for values obtained on average for late-type (small square) and early-type (large square)  SDSS galaxies \citep{2005MNRAS.362...41G}. As previously, we decrease the observed metallicity values. In addition since the SDSS galaxy sample is biased toward lower-density regions than typically used for early-type galaxy studies \citep{2005MNRAS.362...41G}, ages are shifted by the mean age difference between galaxies inside and outside clusters reported above, i.e., 2.1 Gyr. As expected, since late-type galaxies are underrepresented  in the cluster, few simulated galaxies (colored filled small circles) match the small squares with a fraction increasing from inside to outward. On the contrary, most of our galaxies match the larger square although simulated galaxies are slightly older on average despite the shift. Even galaxies in the suburbs are affected by the cluster density. We note, however, that had we  taken 2.5 Gyr ($\sim$0.4~dex) rather than 2.1 Gyr ($\sim$0.32~dex), the shift would be inexistent for most masses except for the low-mass galaxies. We also note  that on the observational side, the age depends on the mix of star formation histories in the model and can vary by up to 0.07~dex depending on the model used \citep{2005MNRAS.362...41G}. The residual shift is thus rather insignificant. Moreover, early-type galaxies from Atlas$^{3D}$ \citep{2015MNRAS.448.3484M}, represented by gray filled small squares without age shift, that are partly within the Virgo cluster confirm that  ages are equivalent in the simulation and in the observations. We note that, in addition, no metallicity shift is applied to  Atlas$^{3D}$ because the aperture is larger (85\% of the light) than that used by \citet[30\% of the light]{2005MNRAS.362...41G}. Measuring the metallicity within one-eighth of the effective radius, half this radius, or this radius  \citet{2015MNRAS.448.3484M} showed that metallicities are offset by about --0.2-0.3~dex. Overall, this means that the value difference shown in Fig. \ref{fig:stellarmetal} between our simulated metallicities and those observed is probably due to a combination of both observational aperture and numerical modeling effects. Thus, the difference due to numerical modeling is not as important as it first appeared.

Our findings confirm that galaxies in clusters are older, and thus more metal rich on average, and that even those beyond the virial radius are affected by this dense environment up to several times the virial radius. Additionally, it seems that the slope of the age--metallicity relation becomes steeper with increasing density of the region. It is actually steeper than in the lower-density regions. From right to left, the slope rises more clearly when considering galaxies by mass bin. The increasing mass--metallicity relation visible in  Fig. \ref{fig:stellarmetal} is on average recovered, except for the high-mass end mostly because of low statistics. This rising slope with increasing density suggests that a closed-box scenario is not valid. A variety of gas accretion and ejection histories linked to the environment is required. In other words, {a low metallicity does not necessarily imply a young stellar population,} but more likely quenching especially in a dense environment. The environmental factor takes precedence over the age.

To summarize compared with previous simulations using the AMR \ramses\ code \citep[e.g.,][]{2016MNRAS.459.4408M}, although our metallicities are still slightly too low, the transition regime is visible in the mass--metallicity relation. The AGN prescription quenches  high-mass galaxies more effectively.  Regarding the metallicity, we propose   investigating variations in the yield. It could indeed be different in massive clusters than in the field \citep{2014MNRAS.444.3581R}.  The deficit in metallicity might still be due to a lack of galactic winds following supernova explosions because of a lack of resolution \citep{2011MNRAS.417.1853D}, but mostly the mass returned and yield need increasing. Investigating IMF with different efficiencies is the first path we will follow in a future work with the next running simulation. 

 \subsection{{Cosmic time: Quenching precedes infall for massive and low-mass galaxies}}
 
  \begin{figure*}
  \vspace{-0.5cm}
  
  \centering
\includegraphics[width=0.8 \textwidth]{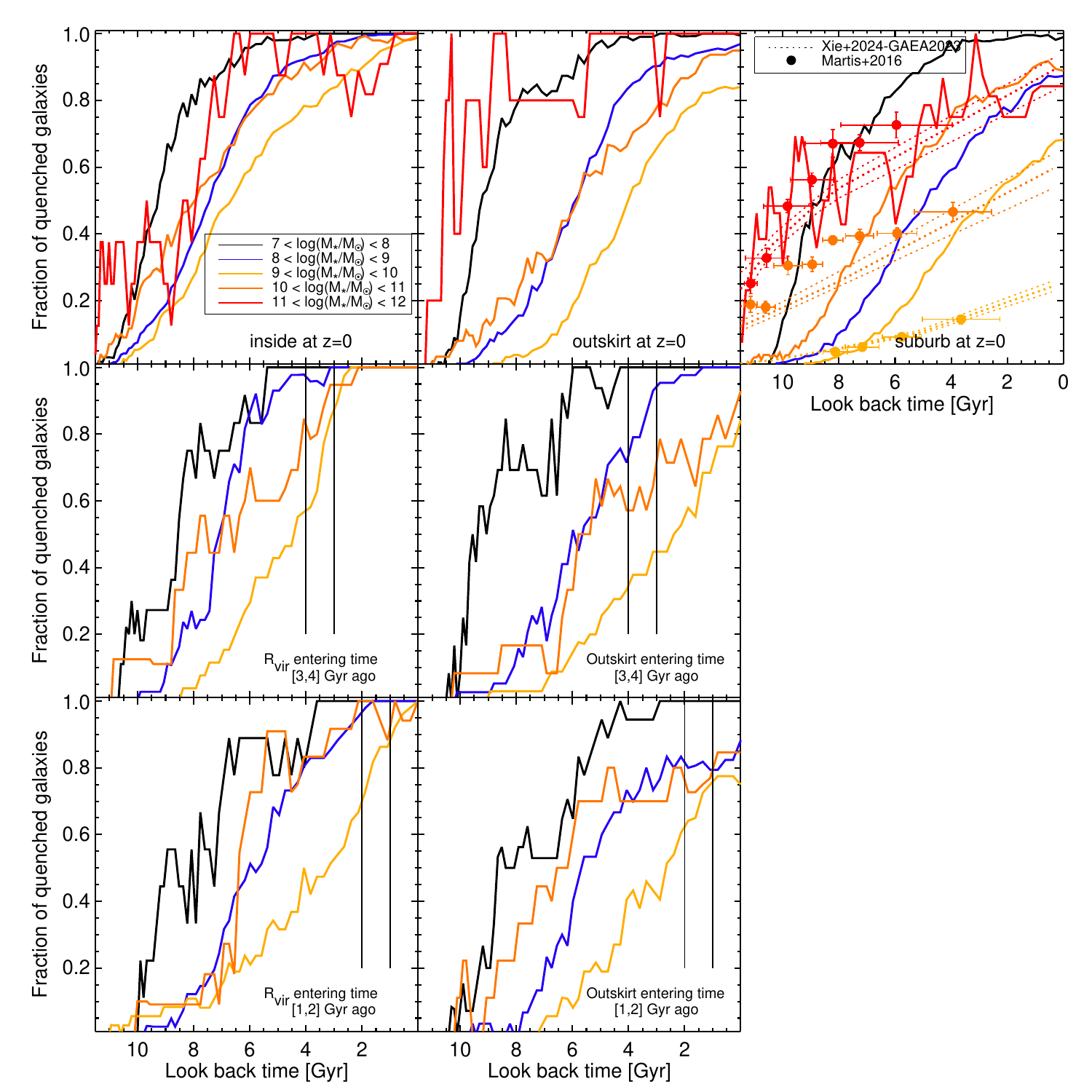}\\
\caption{\textit{Top (from left to right)}: Evolution of the fraction of quenched galaxies across cosmic times for galaxies inside the cluster, in its outskirts, and in its suburbs at $z=0$. In addition, in the last panel, observational data from the UltraVISTA DR1 and 3D-HST surveys, regardless of the environment \citep{2016ApJ...827L..25M}, are shown as filled circles and error bars using similar bins of mass and color-coding. The dotted lines using the same color-coding stand for data obtained with the GAEA2023 semi-analytical modeling \citep{2024ApJ...966L...2X}. \textit{Middle}: Same, but for galaxies that entered the virial radius and the outskirts between 3 and 4 Gyr ago. \textit{Bottom}: Same, but with entering time between 1 and 2 Gyr ago. The colors stand for the different mass bins. Three-dimensional distances are used. The vertical solid black lines delimit the entering time of the galaxies; the quenched fractions are derived, within either the virial radius or the outskirts. While the probability of being quenched for a galaxy depends on its mass when still relatively far from the cluster, this probability rapidly evolves once entering the virial radius to become cluster-mass dependent.}
\label{fig:evotime}
\end{figure*}

 Satellite quenching is thought to occur several gigayears after infall in the cluster \citep[e.g.,][]{2012MNRAS.424..232W}. It is thus of interest to look at the evolution of galaxy properties and more generally of their quenched fraction as a function of cosmic time.  The top panel of Fig. \ref{fig:evotime} shows the evolution of the fraction of quenched galaxies binned by mass across cosmic times, be they at $z=0$ inside the cluster or in its outskirts or suburbs (from left to right). Quenching starts to occur early on. Intermediate-mass galaxies are clearly those that are not affected for the longest time as their quenched fraction is always the lowest. The quenched fraction for the smallest galaxies seems to evolve at the same speed whatever the distance to the cluster. Clearly, with their low gravitational potential, they are most probably very efficiently affected by ram-pressure stripping even at large distances, although internal supernova feedback processes cannot be discarded. On the contrary intermediate-mass galaxies and even the largest one are less likely to be quenched with  increasing distance from the cluster.  In the last panel of the top row of  Fig. \ref{fig:evotime}, the observational data from the UltraVISTA DR1 and 3D-HST surveys, i.e., from whatever environment  \citep[filled circles and error bars,][]{2016ApJ...827L..25M}, confirm that the highest stellar mass galaxies are not affected by the cluster environment when in the suburb area at $z=0$. They need to be deeper into the cluster to be affected. The red filled observational circles overlap the solid red line obtained with our synthetic population, and both are consistent with the GAEA2023 modeling  \citep[dotted lines,][]{2024ApJ...966L...2X}. Conversely, lower stellar mass galaxies are already affected in the suburbs of the cluster. Hence, the quenched fractions for the lower-mass bins increase faster in the suburbs of our simulated cluster than in a general galaxy population such as the synthetic galaxy sample of \citet{2024ApJ...966L...2X}  and the observed galaxy sample of \citet{2016ApJ...827L..25M}.  
 
 While a long time ago, each mass bin had its own fraction of quenched galaxies; the  differences between the fractions tend to be erased with time in the cluster. In agreement with the observations, which suggest that far away from the cluster the stellar mass is the first indication of the likeliness of being quenched for a galaxy, while in the cluster, the mass of the latter gives the probability for a galaxy to be quenched \citep{2013MNRAS.428.3306W}. This is even more visible when restricting the galaxy sample to those entering the virial radius between 3 and 4 Gyr ago, as shown in the second row of  Fig. \ref{fig:evotime}. While fractions were quite different initially and even quite stable for a few gigayears, they start rising with two distinct trends. Fractions for the smallest and highest-mass galaxies after a fast increase continue rising with a similar shallow slope. On the contrary, the fraction for intermediate-mass galaxies after no rise starts increasing with a steep slope so that fractions can become similar in the cluster.  The same behavior is observed when changing the entry time, as shown in the last row of the same figure. The trend is much less pronounced when considering galaxies entering the outskirts. Notably, the most different behavior is seen for the most massive galaxies, and confirms that contrary to the lower-mass galaxies, massive galaxies need to get closer to the cluster core to start feeling the effect of the cluster environment on their SFR. 
 
 The fact that the quenched fraction does not rise immediately to one after galaxies enter in the virial radius is a comfort to us, in  that quenching is not instantaneous. However, the virial radius seems to be a good proxy for an entry point speeding up the quenched fraction increase in agreement with observations. All these findings are in agreement with observations suggesting that galaxies (lower mass aside) start  experiencing environmental effects once in the virial radius of the cluster or at least in its outskirts, but stay active for several gigayears, typically one orbital time, before  being quenched quickly, i.e., in less than 0.5~Gyr \citep[e.g.,][]{1972ApJ...176....1G,2011MNRAS.413..996M,2013MNRAS.428.3306W}.  
 
%%%%%%%%%%%%%%%%%%%%%%%%%%%%%%%%%%%%%%%%%%%%%%%%%%%%%%%%%%%%%
%%%%%%%%%%%%%%%%%%%%%%%%%%%%%%%%%%%%%%%%%%%%%%%%%%%%%%%%%%%%%
%CONCLUSION%%%%%%%%%%%%%%%%%%%%%%%%%%%%%%%%%%%%%%%%%%%%%%%%%%%%%
%%%%%%%%%%%%%%%%%%%%%%%%%%%%%%%%%%%%%%%%%%%%%%%%%%%%%%%%%%%%%
%%%%%%%%%%%%%%%%%%%%%%%%%%%%%%%%%%%%%%%%%%%%%%%%%%%%%%%%%%%%%

\section{Conclusion}

The Virgo cluster of galaxies is our closest cluster-neighbor. As such, it is a great object of study to understand cluster formation and the evolution of galaxies once they enter this rich environment. For this paper we used the first full zoom-in hydrodynamical simulation of a counterpart of the Virgo cluster of galaxies that was obtained by constraining initial conditions with only local galaxy radial peculiar velocities. The large-scale environment of the cluster resembles our local neighborhood and the simulated cluster shares the same formation history as its observed version. The zoom-in region is $\sim$30~Mpc with an effective resolution of 8192$^3$ particles for the highest level (particle mass of 3$\times$10$^7$~M$_\odot$) and a refinement down to $\sim$350~pc. AGN and supernova feedbacks were included. 

A global study of the simulated galaxy population confirms the agreement with the general observational expectations for galaxy clusters: galaxies can be divided into distinct subpopulations depending on their distance from the cluster center. The reddest and oldest galaxies dominate the core, while bluer and star-forming systems lie in the suburbs. Clearly, proximity to the cluster quenches star formation; galaxies that approach and enter the cluster quickly turn red and passive. 

The star formation density within a $\sim$12 Mpc region centered on the cluster agrees well with that inferred from SDSS data and is only slightly smaller in the cluster region.  The match could appear surprising as the SDSS sample is biased toward field galaxies. Further investigations revealed that the larger number of galaxies in the cluster than in the field overcompensate for the fact that at a given mass galaxies in the cluster indeed form fewer stars.   

Quenched fractions reproduce the observed trends. There is a minimum at intermediate-mass galaxies. The smallest galaxies are pre-processed even before entering the cluster probably as satellites and because gas depletion in particular via ram pressure is very efficient on their shallow potential. The largest galaxies need to get close to the cluster center to feel the environmental effect of the cluster.  {Although our quenched fraction may seem too high, projection effects with the main filament linking the Virgo cluster in the line of sight, absence of observational uncertainties and single value  (not an average over several cluster values) could explain most of this discrepancy. {Moreover, our simulation manages to have a sufficiently high quenched fraction at the high-mass end with the proper increasing trend with mass}. {We note that the high quenched fractions may also reflect a combination of cluster-specific assembly history and modeling choices, including the adopted IMF and feedback prescriptions. In this sense, numerical over-quenching cannot be excluded. Exploring halo-to-halo variance within constrained simulation suites will be essential in order to assess the generality of these results}.

Regarding metallicity, we recover the mass-metallicity and age-metallicity relation,  although the metallicities are still perhaps slightly too low. However, observations may have their share of responsibilities because of the star formation history modeling, the S/N cutoff, and especially because of the small aperture radius. However, the break at intermediate mass and the scatter of the mass-metallicity relation are recovered. The age-metallicity slope appears to get steeper when the galaxies are closer to the cluster center. {Low metallicity no longer indicates youth there, but reflects quenching. Overall, galaxies in the cluster are older than their field counterparts, even when compared with field early types.}

 {To summarize, at z = 0, we find a reduced SFR at fixed stellar mass, leading to a flatter SFR–stellar mass relation than at higher redshift. The probability of quenching is stronger for low-mass galaxies in the inner core, while intermediate-mass or isolated galaxies tend to remain only partially quenched. Quenched systems also experienced significant dark matter stripping, and gas depletion is pervasive for low-mass galaxies out to three virial radii, showing that environmental influence extends far beyond the virial boundary.}

In addition, a study across cosmic time reveals another perfect agreement with observations: quenching outside of the cluster depends mostly on the galaxy mass, while in the cluster it seems to depend on the cluster mass. Additionally, quenching is more likely for satellite galaxies. {In general, quenching is found to precede infall for both massive and low-mass galaxies.}

In a previous paper, general comparisons between the observed population of the Virgo cluster and the simulated population revealed excellent agreement in terms of luminosity and mass distribution as a function of the cluster center and the presence of a M87 counterpart. Additionally, the history of the simulated cluster is remarkably similar to the observed one: 1) about 200--300 small galaxies entered the cluster within the last 500-1000~Myr ; 2) the latest big merger started 3 to 4~Gyr ago, it was a group that finished merging within the last gigayear or so. It was about 10\% the mass of the cluster today and it entered the cluster via the filament diametrically opposed to us.

In  summary, this excellent numerical replica of the Virgo cluster of galaxies combined with an efficient hydrodynamical modeling will permit a study of the different galaxy populations (e.g., jellyfish, backsplash, the fall-in small group, M87) in the simulated cluster in detail and will allow us to compare them directly with their real counterparts within the observed cluster. The hot gas phase of the simulated cluster will also be part of a future study. Meanwhile another series of four papers is studying the dynamics and structure of the replica's intracluster medium to show how gas motions, turbulence, and projection effects impact observable properties. Since Virgo is known to be a cool-core cluster and this property is most probably due to its history, it will be an excellent way of testing baryonic physics in the intracluster medium. Further investigations will focus on the low metallicity obtained in our simulated cluster galaxies as well as the perhaps too high quenched galaxy fractions. Our first attempt is a currently running simulation using a Kroupa IMF-based metal yield and mass return time. This simulated Virgo cluster, in good agreement with observations tested to date, opens great prospectives and lets us foresee  multiple interesting projects.  

\begin{acknowledgements}
The authors would like to thank L\'eo Michel-Dansac for useful discussions, {and the referee for their comments}. This research has made use of the SIMBAD database, operated at CDS, Strasbourg, France as well as of the Extragalactic Distance Database (http://edd.ifa.hawaii.edu). The authors acknowledge the Gauss Centre for Supercomputing e.V. (www.gauss-centre.eu) for funding this project by providing computing time on the GCS Supercomputer SuperMUC-NG at Leibniz Supercomputing Centre (www.lrz.de), project ID:~pr74je. This work was supported by the ``action th\'ematique'' Cosmology-Galaxies (ATCG) of the CNRS/INSU PN Astro. JS is supported by the University of Lille via the Welcoming Internationals to Lille initiative for the UNIVERSITWINS project. AK and GY are partially supported by the Spanish Ministerio de Ciencia e Innovación, (MICINN) under research grant PID2021-122603NB-C21 as well as project PID2024-156100NB-C21 financed by MICIU /AEI/10.13039/501100011033/FEDER, UE. AK further thanks Pacific for barnoon hill.
\end{acknowledgements}

%%%%%%%%%%%%%%%%%%%%%%%%%%%%%%%%%%%%%%%%%%%%%%%%%%%%%%%%%%%%%%%%
\bibliographystyle{aa}

\bibliography{biblicompletenew}

%%%%%%%%%%%%%%% APPENDIX %%%%%%%%%%%%%%%%%%%
\begin{appendix}

 \section{{Mass dependence of galaxy properties within the cluster-centric distance at $z=0$}}
 \label{appendixA}
 
  \begin{figure*}
\vspace{-1.5cm}
\centering 

\includegraphics[width=0.47 \textwidth]{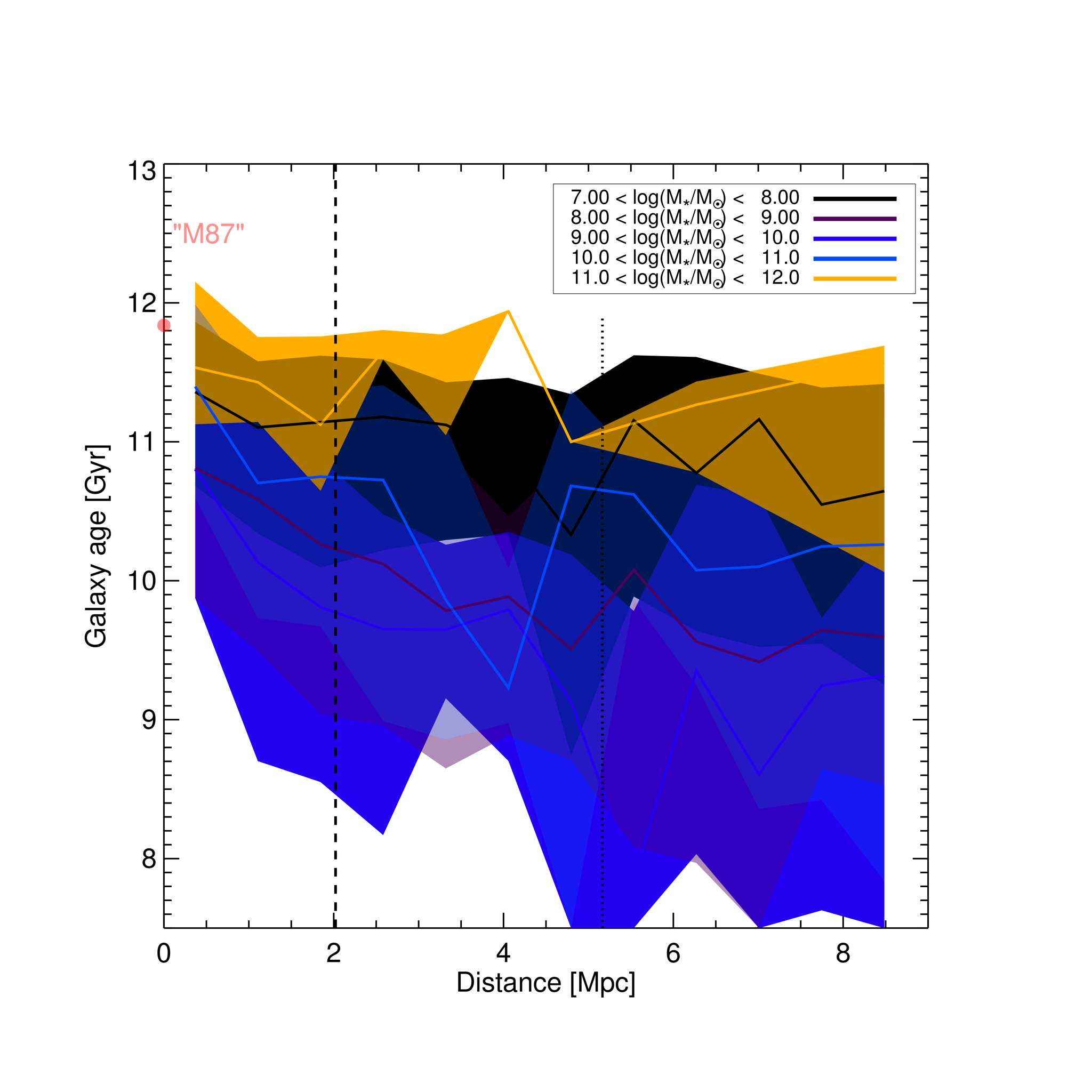}  
\includegraphics[width=0.47 \textwidth]{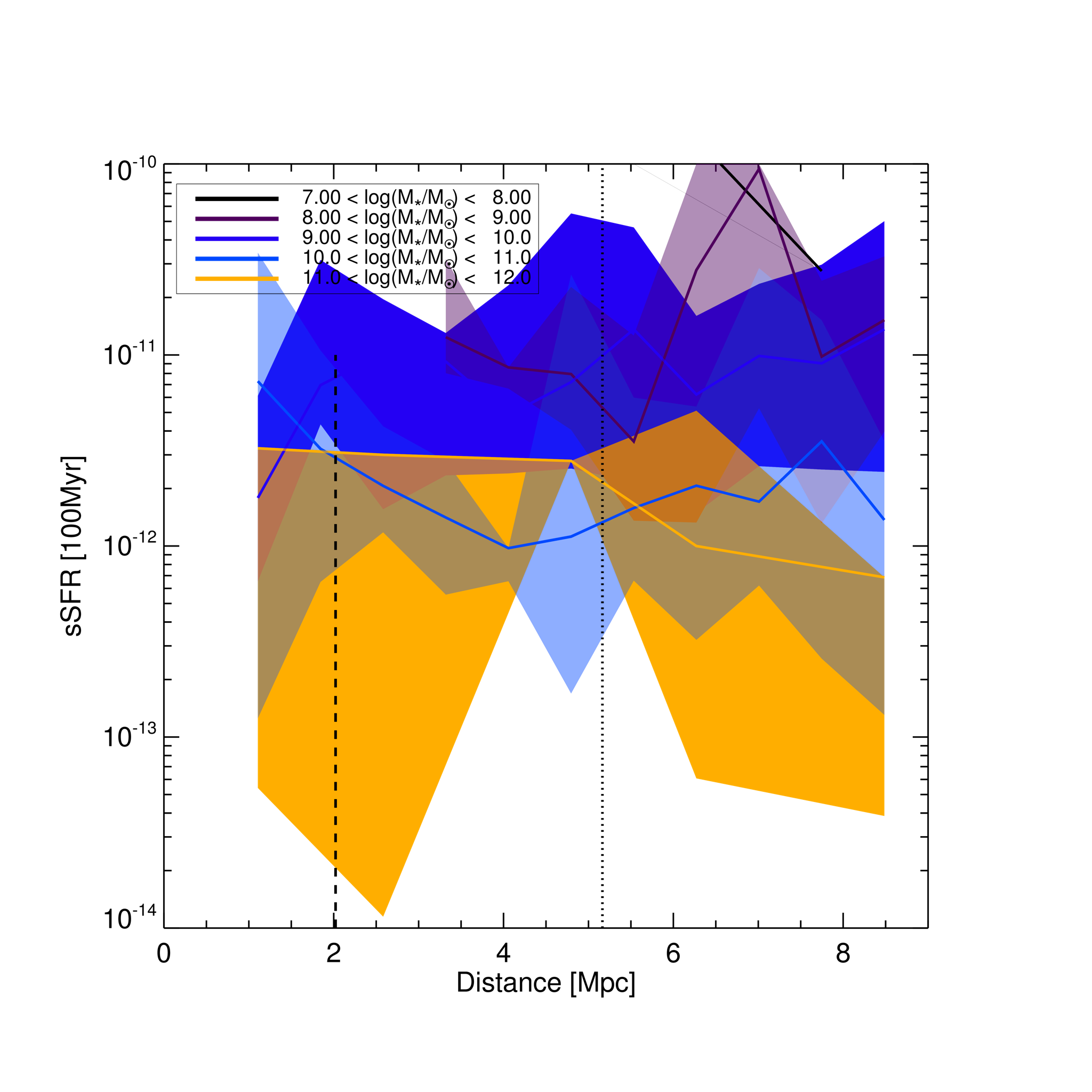}\\ 
\vspace{-2.cm}

\includegraphics[width=0.47 \textwidth]{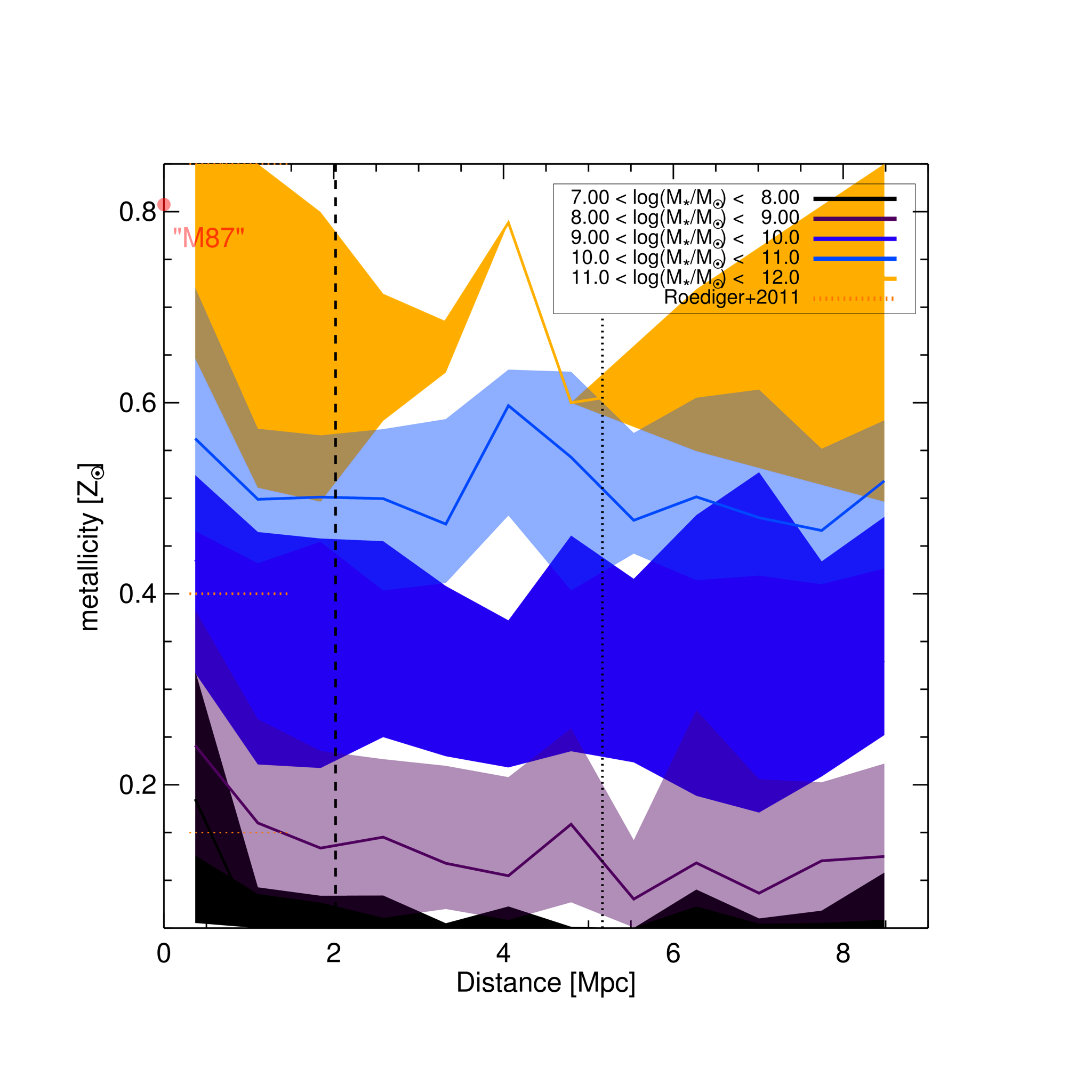} 
\includegraphics[width=0.47 \textwidth]{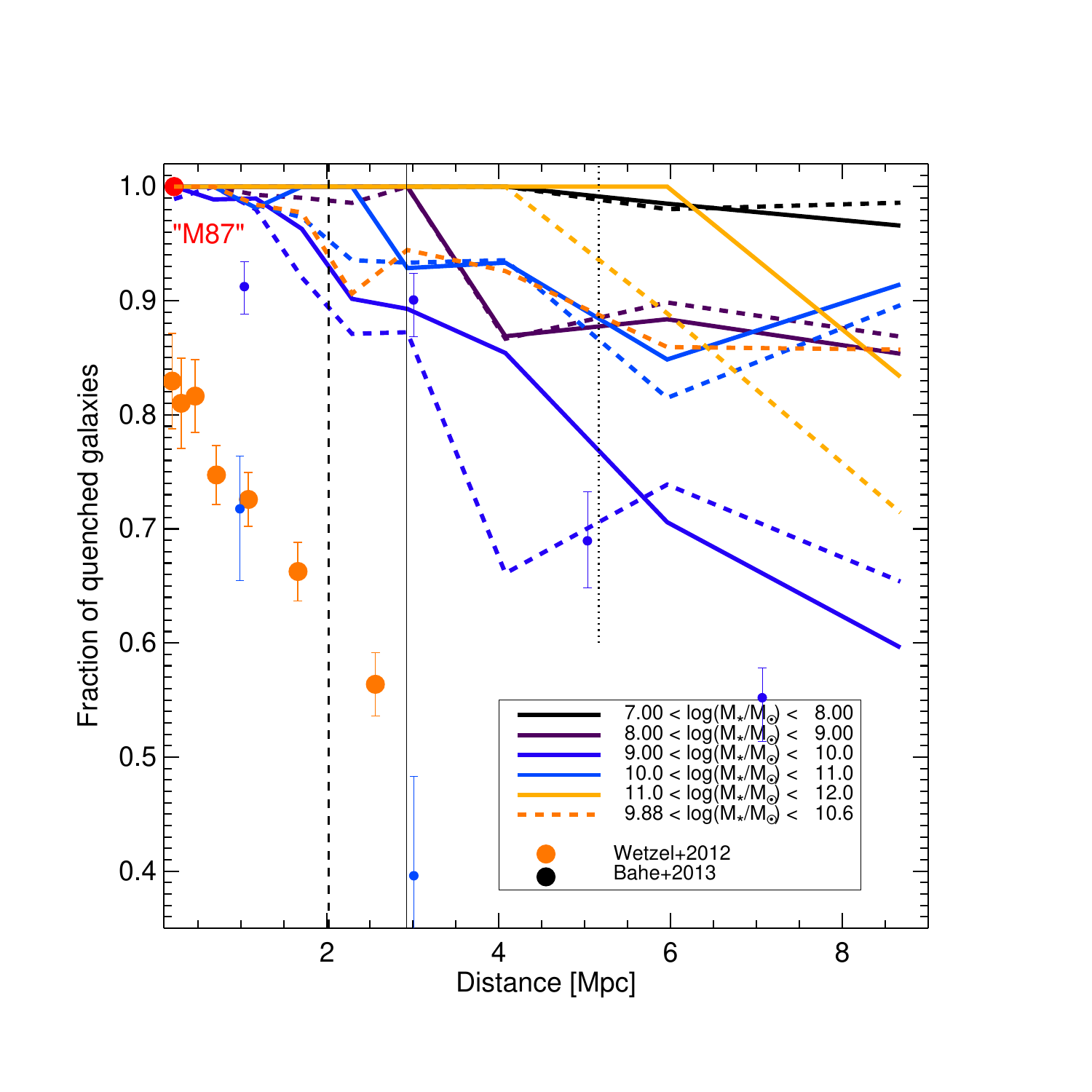}\\
\vspace{-0.7cm}

\caption{Median properties of galaxies per bin of mass as a function of  distance to the cluster center. \textit{From left to right, top to bottom}: Age, sSFR over 100~Myr, metallicity, and fraction of quenched galaxies. From black to yellow, galaxies are heavier and heavier. The red filled circles stands for the simulated counterpart of M87. Dashed and dotted black lines represent the virial and zero velocity radii. Solid lines give median values per distance bin and transparent areas stand for the 16th and 84th percentiles. \textit{Bottom left}: Short dotted  orange lines stand for the average and standard deviations of metallicities measured by \citet{2011MNRAS.416.1996R} in the inner part of the Virgo cluster. \textit{Bottom right}: Orange filled circles with error bars are observationally derived average fractions by \citet{2012MNRAS.424..232W} and blue filled circles are obtained from the simulations of \citet{bahe13}. The color-coding refers to the mass range similarly to that used for our synthetic galaxies. The dashed lines are based on projected rather than 3D distances in a $\sim$12~Mpc radius volume. Galaxy properties tend to evolve with the distance from the cluster center: their age and likeliness of being quenched decrease with the distance or conversely their sSFR increases with intermediate-mass galaxies presenting the largest gradient.}
\label{fig:evodistance}
\end{figure*}

It is of great interest to look at the properties of the galaxies as a function of the distance to the cluster center in more depth. Galaxy properties are indeed known to correlate with the cluster-centric distance even within the virial radius. Hence, the galaxy population changes with the distance to the cluster center \citep{1980ApJS...42..565D}. The segregation of galaxies as a function of the cluster-centric distance has been largely quantified with splitting galaxies according to different properties \citep[e.g.,][]{1993ApJ...407..489W,2001AJ....121.1266D,2002A&A...387....8B} and verified numerically at the high-mass end  \citep{2016MNRAS.459.4408M}. In this appendix we thus extend the analysis of \citet{2016MNRAS.459.4408M} made with Rhapsody simulations \citep{2017MNRAS.470..166H} to lower-mass galaxies.  

The youngest galaxies are those of intermediate size\footnote{This is roughly the Milky-Way size.} and they are not within the cluster core. We recover that galaxies in the cluster tend to be on average older than those outside whatever their mass is. Galaxies in the cluster are quenched or alternatively they have a low, if not null especially for the lowest-mass galaxies, sSFR whatever their mass. Star-forming galaxies indeed tend to avoid the cluster center \citep{2003AJ....126.2662D}\footnote{{Galaxy morphology is indeed known to correlate strongly with star formation activity, with late-type galaxies typically being star-forming and early-type systems predominantly quiescent.}}. The sSFR starts to increase, or be not null, in the outskirts, again more specifically for the lowest-mass galaxies, confirming that the smallest galaxies undergo a transformation before entering the virial radius of the cluster. Galaxy metallicity is quasi constant with the cluster-centric distance, and with values in agreement with results from \citet{2011MNRAS.416.1996R} shown with the short dotted orange lines in the cluster core. The correlation mass - metallicity is recovered: the more massive a galaxy is the more metal rich it is. We note that \citet{2002AJ....123.1807G} showed that all the galaxy population segregations break at distances larger than the virial radius, here we observe  breaks mostly in the outskirts, specifically between the virial and zero-velocity radii.

Regarding the quenched fraction of galaxies as a function of the clustercentric distance, for example \citet{2019MNRAS.483.3336T} in its ROMULUSC simulation, we find that our results are biased high compared  ({+25\%}) to observational expectations obtained by \citet{2012MNRAS.424..232W} and shown as filled orange circles in the last panel of  Fig. \ref{fig:evodistance}. Our projected distances, using the same mass bin as theirs (dotted orange line), are limited to a $\sim$12~Mpc radius sphere against their ten times the virial radius but it is doubtable that even with a region as large as theirs, the same fraction can be recovered within the simulation, even though we are still biased high by the main filament linking the Virgo cluster to the cosmic web in our line of sight. However, they also combined results from several halos with masses between 10$^{14.1}$ and 10$^{15}$~M$_\odot$. Their sample is thus biased toward the low-mass end compared to our simulated cluster. Since the quenched fraction increase with the cluster mass, this is another argument in favor of a decreasing discrepancy between our simulated quenched fraction and the observed values. Finally, our low statistics also prevent us from using the same small distance bin. It is thus extremely difficult to judge up to which amount our galaxies are too quenched. Interestingly,  \citet{bahe13} found the same quenched fraction as theirs (filled blue circles) but for a slightly different bin of galaxy stellar mass and without using projected distances. This bin of stellar mass corresponds also to their bin of stellar mass with the minimum quenching fraction while ours is at a smaller stellar mass bin. This confirms the slight shift in stellar mass obtained for the minimal quenched fraction observed in  Fig. \ref{fig:quenched} that could be related to our IMF choice even if values are rescaled. More importantly, we recover the proper trend.

In addition, according to \citet{2013MNRAS.428.3306W}, quenching is a function of stellar mass mostly far from the cluster. Close by, the host mass seems more important \citep[see also][]{2012MNRAS.424..232W}. One simulated cluster does not allow us to thoroughly verify this; however, it does seem that the farther away from the cluster galaxies are, the more distinguishable their quenched fractions are given their stellar mass. Regardless, this trend is consistent with quenching being also nurtured by the environmental density rather than being purely natural. Phenomena such as shock heating, strangulation, tidal stripping, and ram-pressure stripping most certainly come into play still differently according to the galaxy stellar mass (see also our discussion in the core of the paper). It will be interesting to study more thoroughly  in the future at what level and to what amount these different mechanisms happen.

\section{Additional figure}
\label{appendixB}

{This appendix shows the same figure as Fig. \ref{fig:stellarssfr} but for the sSFR.} The thin three dots-dashed yellow lines show the scatter of the relation found by \citet{brinchmann04}. 

\begin{figure}[H]
\vspace{-0.9cm}

\centering
\includegraphics[width=0.47\textwidth]{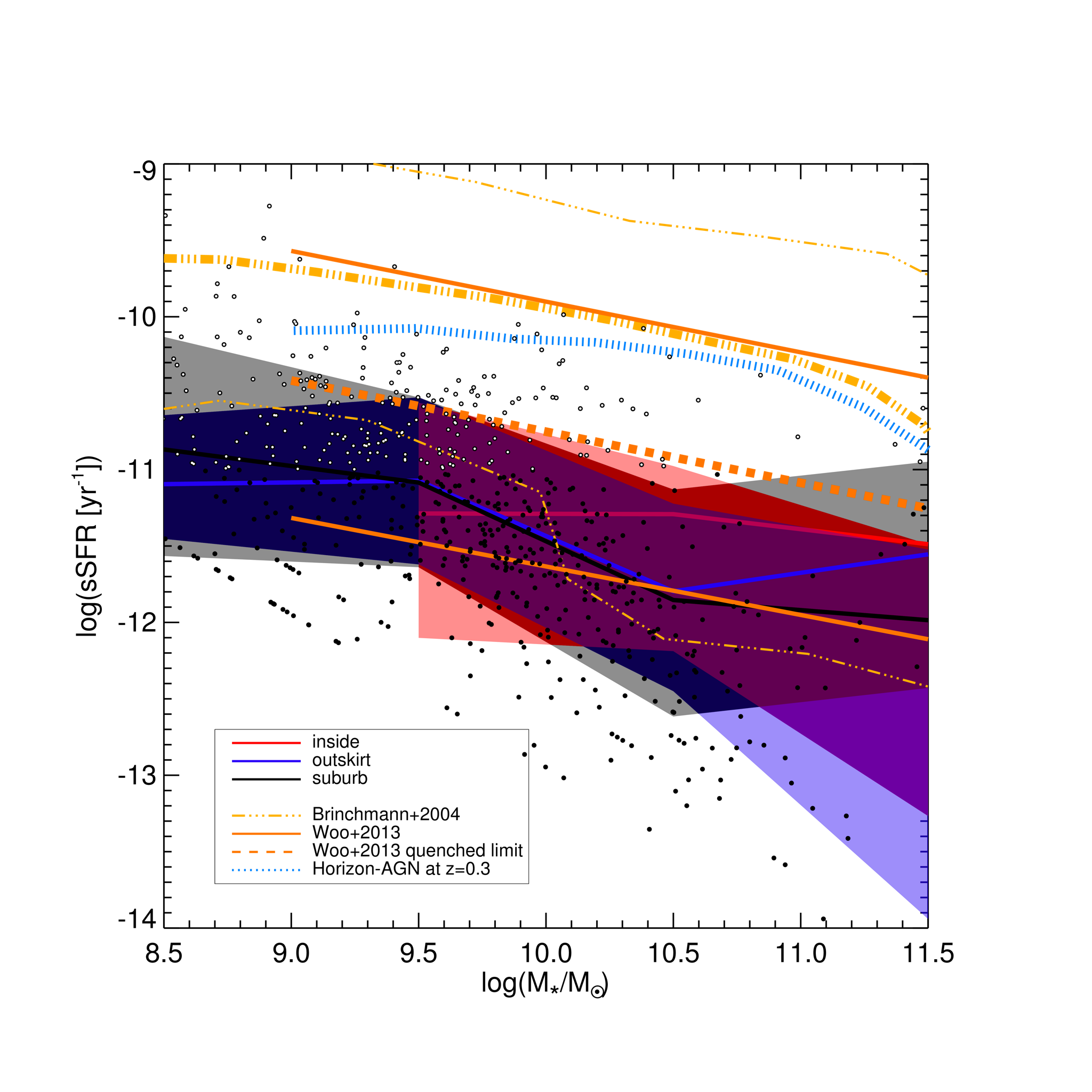}\\
\vspace{-0.4cm}

\caption{Same as Fig. \ref{fig:stellarssfr} but for the sSFR. In addition, the thin three-dot-dashed yellow lines show the scatter for the observed galaxies. Moreover, the dotted light blue line gives the relation found for the Horizon-AGN galaxies at $z=0.3$ \citep{2016MNRAS.463.3948D}. Most simulated galaxies are quenched, meaning that the cluster rich environment affects galaxy star formation out to large radii.}
\label{fig:ssfr}
\end{figure}

\section{Standard error on the median}
\label{appendixC}

This appendix presents the same figures as in the core of the paper but replaces the percentiles for galaxies inside the cluster, in its outskirts, and in its suburbs by the standard error on the median.

\begin{figure*}
\vspace{-0.9cm}

\centering
\includegraphics[width=0.4\textwidth]{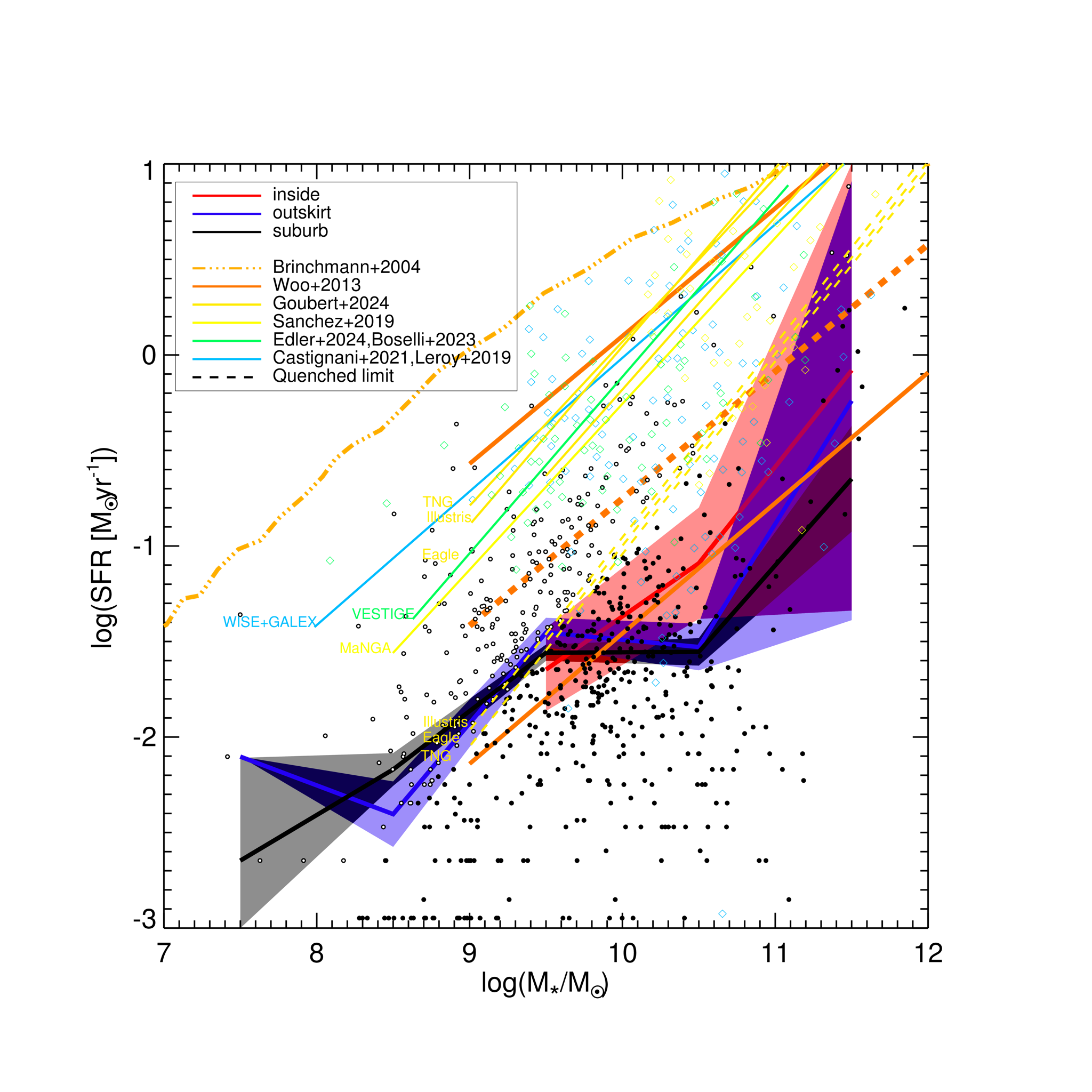}
\includegraphics[width=0.4\textwidth]{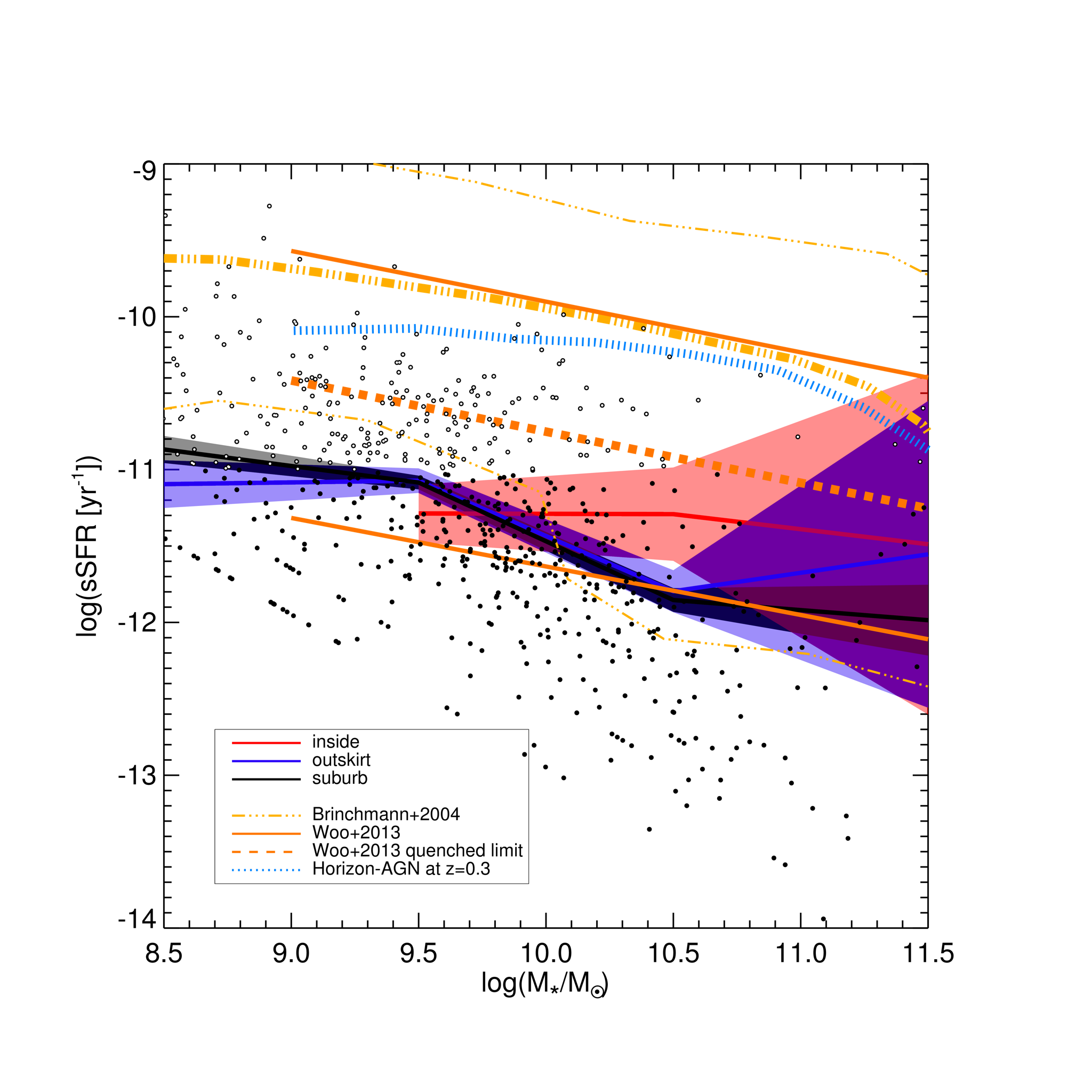}\\
\vspace{-0.7cm}

\includegraphics[width=0.4\textwidth]{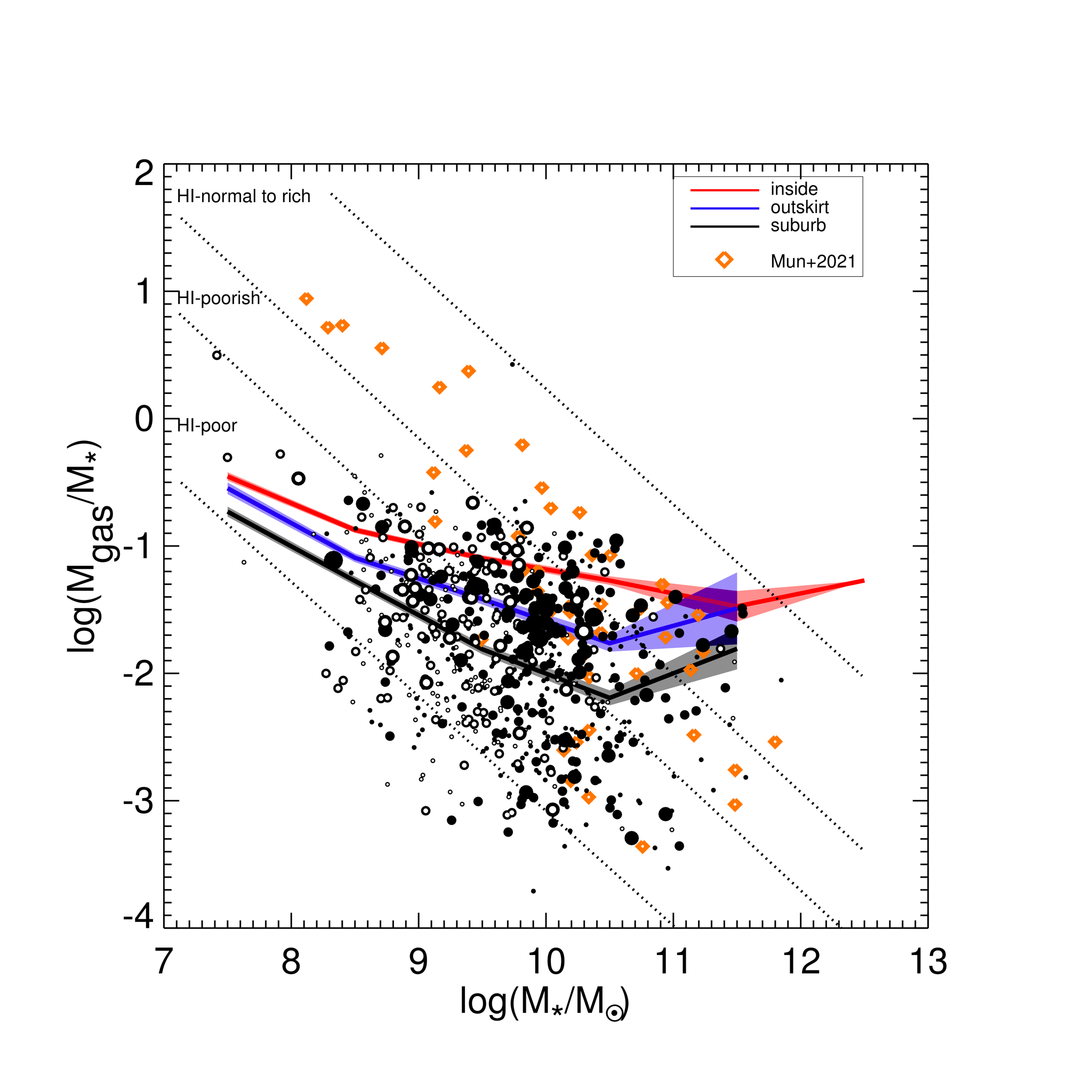}
\includegraphics[trim=0cm 0cm 0cm 10.8cm, clip, width=0.4\textwidth]{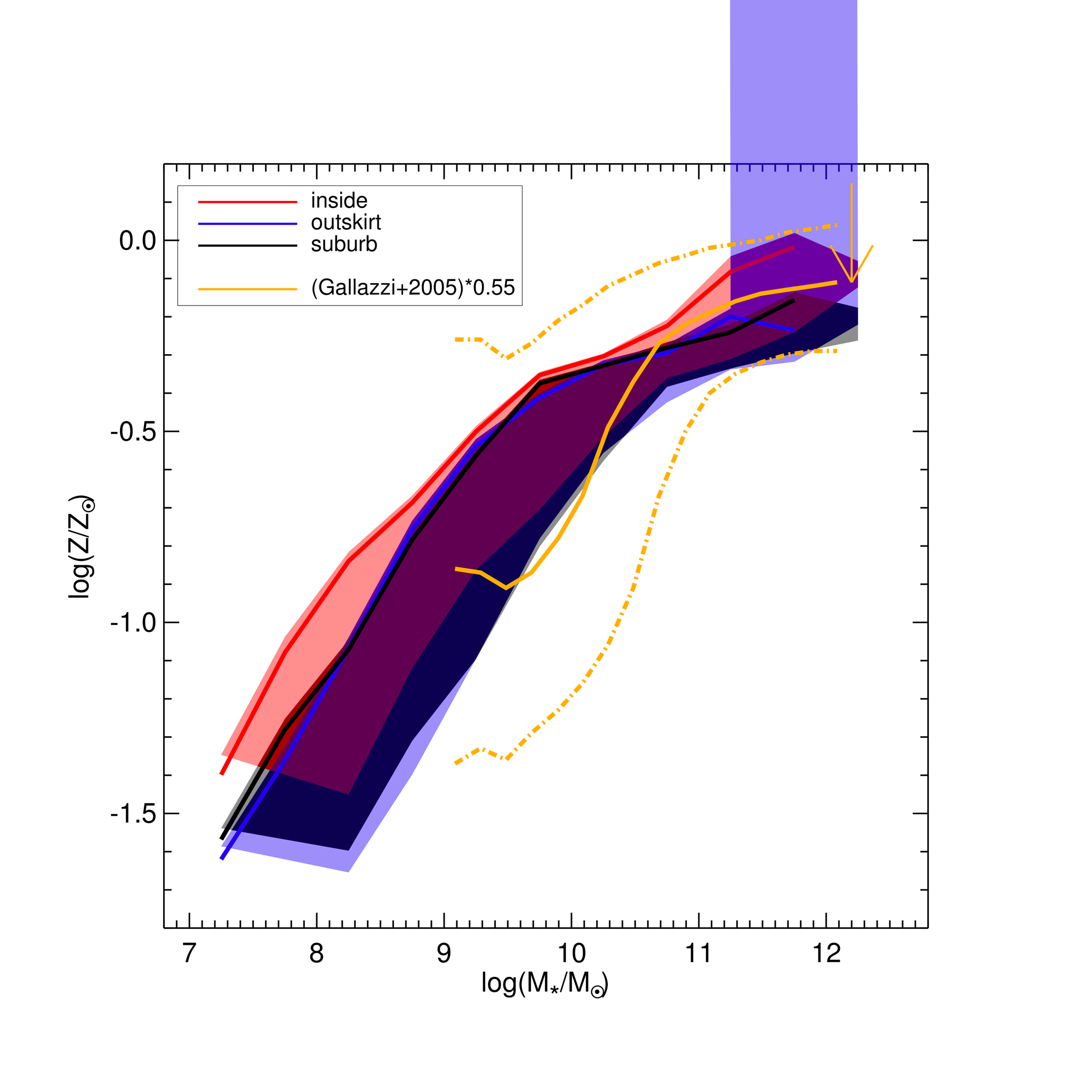}\\

\vspace{0.cm}

\caption{Same as previous figures, but collapsed into a single panel: Fig. \ref{fig:stellarssfr} (top left), Fig. \ref{fig:ssfr} (top right), Fig. \ref{fig:mgasmstar} (bottom left), and Fig. \ref{fig:stellarmetal} (bottom right), except that the transparent areas represent the standard error on the median.}
\label{fig:stellarssfrerrormedian}
\end{figure*}

  \begin{figure*}
\vspace{.5cm}
\centering 

\includegraphics[width=0.35 \textwidth]{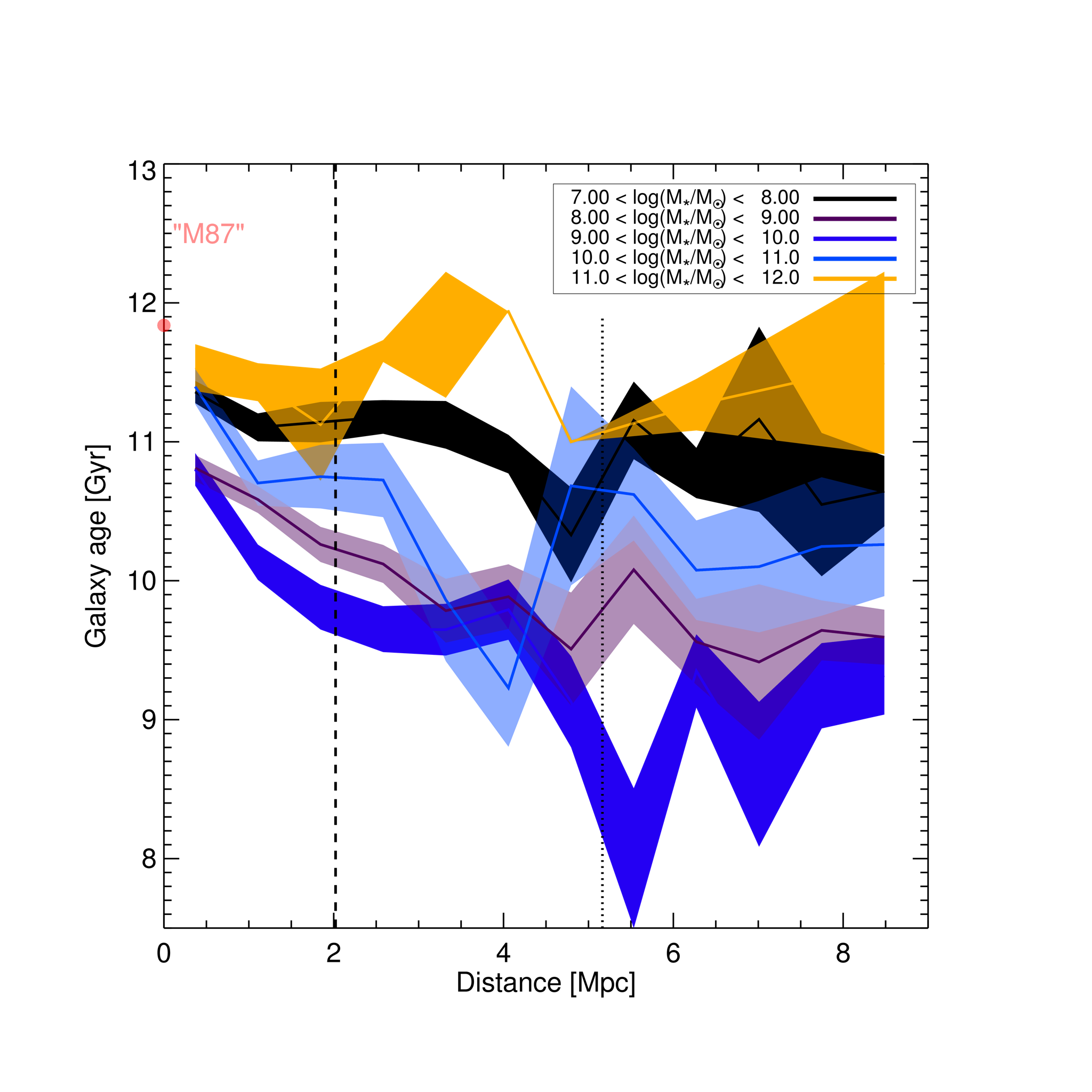}  
\hspace{-1cm}\includegraphics[width=0.35 \textwidth]{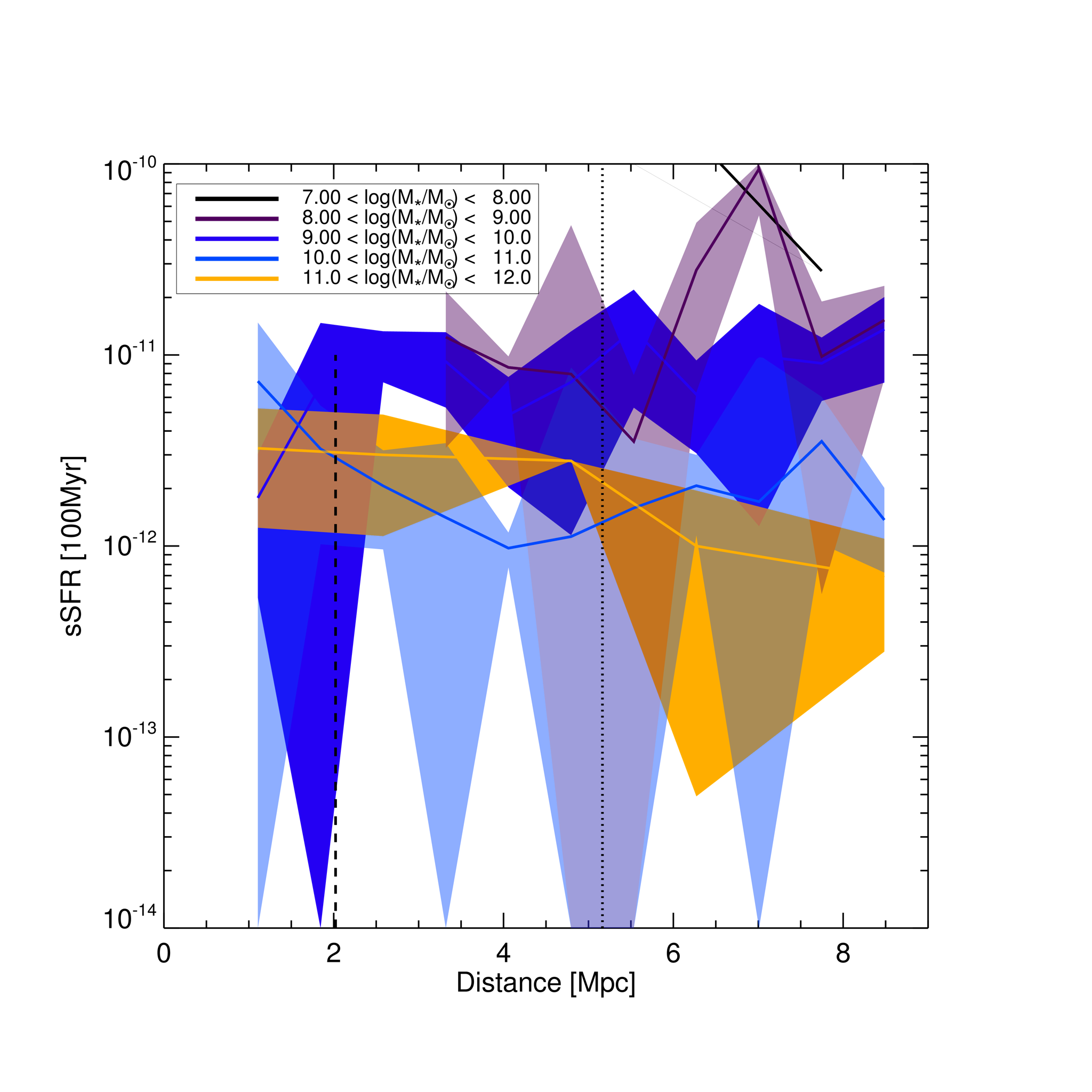}
\hspace{-1cm}\includegraphics[width=0.35 \textwidth]{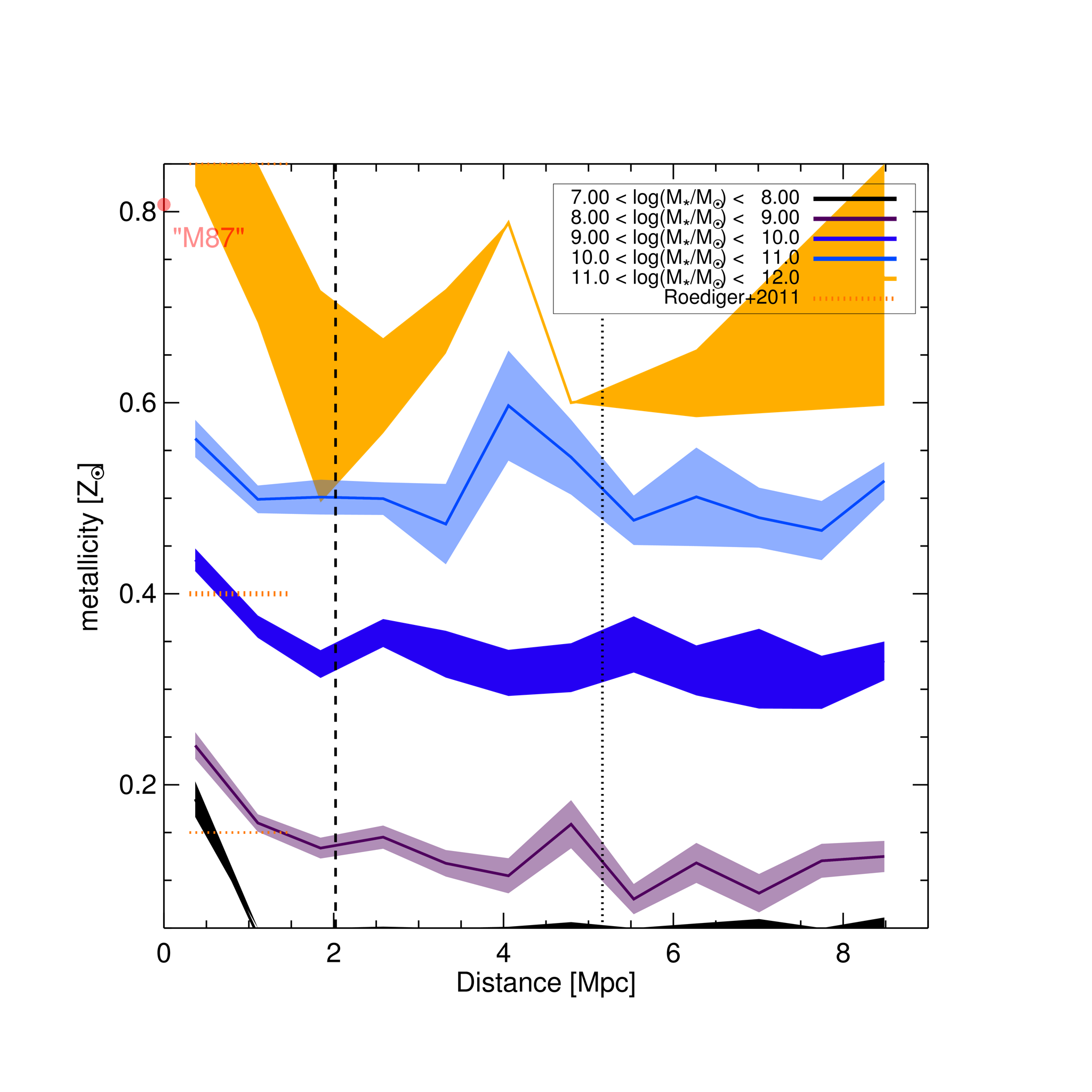} 
\vspace{-0.7cm}

\caption{Same as Fig. \ref{fig:evodistance} (without  the bottom right panel) except that the transparent areas represent the standard error on the median.}
\label{fig:evodistanceerrormedian}
\end{figure*}

\end{appendix}

 \label{lastpage}
\end{document}